\documentclass[useAMS,usenatbib,a4paper,twoside,final]{mn2e}
\usepackage[]{amsmath}
\usepackage{graphicx}
\title[Star cluster evolution in barred disc galaxies]{Star cluster
 evolution in barred disc galaxies. I. Planar periodic orbits}
\author[I. Berentzen and E. Athanassoula]{
 I.~Berentzen~$^{1,2,3}$\thanks{E-mail: iberent@ari.uni-heidelberg.de (IB)}
  and
 E.~Athanassoula~$^{3}$\thanks{E-mail: lia@oamp.fr (EA)} \\
 $^{1}$Astronomisches Rechen-Institut, Zentrum f\"ur Astronomie, Universit\"at Heidelberg, M\"onchhofstra\ss e 12-14, 69120 Heidelberg, Germany\\
 $^{2}$Heidelberg Institute for Theoretical Studies, Schlosswolfsbrunnenweg 35, 69118 Heidelberg, Germany\\ 
 $^{3}$Laboratoire d'Astrophysique de Marseille (LAM), UMR6110, CNRS/Universit{\'e} de Provence,
 Technop{\^o}le de l'Etoile, 38 rue Fr{\'e}d{\`e}ric \\ Joliot-Curie, 13388 Marseille C{\'e}dex 20, France
}
\begin{document}
\date{Accepted 2011 October 6. Received 2011 October 6; in original form 2011
September 4}
\pagerange{\pageref{firstpage}--\pageref{lastpage}} \pubyear{2011}
\label{firstpage}
\maketitle
\begin{abstract}
 The dynamical evolution of stellar clusters is driven to a large
 extent by their environment. Several studies so far have
 considered the effect of tidal fields and their variations, such 
 as, e.g., from giant molecular clouds, galactic discs, or spiral
 arms. In this paper we will concentrate on a tidal field whose effects
 on star clusters have not yet been studied, namely that of bars.
 We present a set of direct $N$-body simulations of star clusters moving in
 an analytic potential representing a barred galaxy. We compare
 the evolution of the clusters moving both on different planar
 periodic orbits in the barred potential and on circular orbits
 in a potential obtained by axisymmetrising its mass distribution.
 We show that both the shape of the underlying orbit and its
 stability have strong impact on the cluster evolution as well as
 the morphology and orientation of the tidal tails and the sub-structures
 therein. We find that the dissolution time-scale of the cluster in
 our simulations is mainly determined by the tidal forcing along
 the orbit and, for a given tidal forcing, only very little by the
 exact shape of the gravitational potential in which the cluster is moving. 
\end{abstract}
%
%-------------------------------------------------------------------------
\section[]{Introduction}
%-------------------------------------------------------------------------

 While the internal dynamical evolution of star clusters is mainly driven
 by relaxation processes, the global galactic environment in which they are
 hosted can drive their evolution externally. Different aspects of star cluster
 evolution in the gravitational field of galaxies have been studied numerously
 using numerical simulations. Recent direct $N$-body simulations include
 the effects of the Galactic tidal field \citep{KMH08, EJS09} and dynamical
 friction terms for clusters close to the galactic centre \citep{JKBES11}.
 From such kind of simulations the picture emerged that star clusters steadily
 loose stars through the Lagrangian points in the system and form long tidal
 tails. In axisymmetric potentials the stars in these tails are found to follow
 orbits which can be approximated by epicycles, resulting in a
 cycloidal motion of stars along the tails 
 when seen in the tidal tail coordinate system as introduced by \citet{EJS09}.
 This results in the formation of sub-structures within the tidal tails in
 form of clumps, often referred to as {\em epicyclic over-densities}.
 Numerical simulations
 have also shown that such structures persist also in tails of star clusters
 on non-circular orbits, however, their local density and position with
 respect to the cluster in this case vary with time \citep{JKBES11,KKBH10}.

 A detailed understanding of the formation, morphology and dynamics of tidal
 tails resulting from tidally disrupted dwarf galaxies and star clusters in
 galaxies are of great interest to observations especially in the Milky Way
 \citep[e.g.,][etc.]{OGR01,BEIHW06,Grill09}, but also in other galaxies
 \citep[e.g.,][etc.]{PFFW02, MDPGMPP09, MDGC10}. Tidal streams are of 
 particular interest for tracing the gravitational potential by
 reconstruction the orbits of the clusters, as e.g., for Palomar\,5
 \citep{OGR01}. 
 The underlying assumption for this method is that the tidal tails
 accurately trace the cluster orbit \citep{VIL11}.

 The galactic potentials used in numerical simulations such as the ones mentioned
 above are usually confined to sphericity or axisymmetry for simplicity. However,
 disc galaxies such as the Milky Way show a variety of morphological and dynamical
 sub-structures that are known to affect the evolution of star clusters. E.g., the
 effect of interactions with giant molecular clouds and spiral arm passages on the
 cluster dissolution time-scale has been studied by \cite{GPBAS06} and \cite{GAP07},
 respectively.

 In this work we focus on the evolution of star clusters in the time-dependent
 gravitational potential of barred disc galaxies. Stellar bars are known to be
 the main internal driver of disc galaxy evolution and are strong non-linear
 perturbations to the gravitational potential of disc galaxies, thus
 having strong effects on the underlying orbital structure in both 2-d and
 3-d. We present a set of direct $N$-body simulations of star clusters moving
 on planar periodic orbits in barred potentials and on circular orbits in
 (near-) axisymmetric potentials. We particularly focus on the formation
 and evolution of tidal tails in both scenarios. 

 The paper is structured as follows: In Section~\ref{sec:numerics} we describe
 the numerical methods and initial conditions used in our $N$-body simulations. 
 Section~\ref{sec:results} then describes the results. We give conclusions in
 Section~\ref{sec:conclusions}.

%---------------------------------------------------------------------------
\section[]{Numerical methods and initial conditions} \label{sec:numerics}
%---------------------------------------------------------------------------

\subsection[]{Direct $N$-body code using GPU clusters}

 We use a modified version of the publicly available direct $N$-body code
 $\varphi$-{\sc Grape} \citep{HGMSPB07}. It provides hierarchical particle
 time-steps and a fourth order Hermite integration scheme \citep[e.g.,][]{MA97}.
 Originally, this code is written for high-performance computing clusters
 equipped with the special-purpose
 hardware {\sc Grape-6}  \citep{MFKN03,FMK05} to accelerate the gravitational
 force calculation. It has been used in the past for simulations of
 dense stellar systems such as galactic nuclei
 \citep[e.g.,][]{BMS05, BMSB06, BPBMS09, KBPPRSS09}. In our current version, the code
 has been adjusted to run on Graphic Processing Units ({\sc Gpu}s) using
 the {\sc Sapporo} library \citep{GHP09} which allows to use the {\sc Gpu}
 cards like the special-purpose hardware {\sc Grape}. 
 
 There are alternative direct $N$-body codes freely available with more advanced
 features, such as regularisation methods for an accurate integration of binaries,
 stellar evolutionary tracks for single and binary stars, etc. However, the
 present work is the first to study star cluster evolution in barred galaxies
 and therefore we restrict ourselves to relatively simple models. Simulations
 with higher resolution, particle mass functions and increasingly more input
 physics will be studied in the future.
 
 In order to prevent the formation of stellar binaries or multiples in our
 simulations, we adopt a small gravitational (Plummer) softening in the force
 calculation. As a trade-off we study the evolution of the star clusters in
 the pre core-collapse phase only, where binaries and multiples are of less
 dynamical importance. As mentioned before, we limit ourselves in this work
 to $N$-body models with equal mass particles. This prevents some
 possible dynamical 
 effects such as mass segregation \citep[see, e.g.,][]{KASS07}. We also
 neglect the dynamical friction between the star cluster and the galaxy,
 which generally becomes important close to the galactic centre. We discuss
 the possible effects of this limitation when necessary and leave simulations
 including this extra physics for the future.
 
 In the next subsection we describe the analytic galaxy model that we have
 implemented in the $\varphi$-{\sc Grape} as an external potential and force
 field.

%--------------------------------------------------------------------------------
\subsection[]{Analytic barred galaxy model}
%--------------------------------------------------------------------------------

 We use an analytic three-component galaxy model which is composed of an
 axisymmetric disc, a spherically symmetric bulge/halo and a triaxial bar
 component. The disc potential $\Phi_{\mathrm{D}}$ in our model is represented
 by a three dimensional Miyamoto-Nagai potential \citep{MN75} of the form

\begin{equation}
\label{eq:mn}
 \Phi_{\mathrm{MN}} (x,y,z) = \frac{G M_{0}}{\sqrt{ x^2 + y^2 + \left(A +
\sqrt{B^2 + z^2}\right)^2}} \ ,
\end{equation}

 \noindent where $G$ is the gravitational constant, $M_{0}$ is the total
 mass of the component, and $A$ and $B$ are the radial and vertical scale-length,
 respectively.
 
 The spherically symmetric bulge/halo potential $\Phi_{\mathrm B}$ of
 our galaxy model 
 is represented by a Plummer potential \citep{Plu11}, i.e., also described by
 Eq.~\ref{eq:mn} with $A=0$.
 For the bar potential $\Phi_{\mathrm{bar}}$ we use a  three-dimensional 
 Ferrers' ellipsoidal potential \citep{Fer77} which is based on a density
 distribution of the form

\begin{eqnarray}
 \rho(x,y,z) 
 = \left\{
   \begin{array}{cl}
     \frac{105\, M_\mathrm{bar}}{32 \pi\, a_{\mathrm{bar}}\, b_{\mathrm{bar}}\, c_{\mathrm{bar}}} \, (1 - m^{2})^{2} & \mbox{ if $m \leq 1$ } \\
     0 & \mbox{ if $m > 1$ } 
  \end{array} \right.   
\end{eqnarray}

\noindent with 
 
\begin{equation}
 \label{eqn:bar}
 m^{2} = \frac{x^{2}}{a_{\mathrm{bar}}^{2}} + \frac{y^{2}}{b_{\mathrm{bar}}^{2}} + 
         \frac{z^{2}}{c_{\mathrm{bar}}^{2}} \mbox{ \hspace{2em} and }
  a_{\mathrm{bar}}>b_{\mathrm{bar}}>c_{\mathrm{bar}} \ ,
\end{equation}
 
\noindent where $M_\mathrm{bar}$ is the mass of the bar component and
  $a_{\mathrm{bar}}$, $b_{\mathrm{bar}}$ and $c_{\mathrm{bar}}$ are its
 semi-principal axes. Note that in our description the bar major axis
 is aligned with the $x$-axis of our coordinate system. The full
 analytic expressions of the gravitational potential $\Phi_{\mathrm{bar}}$
 and corresponding forces can be found in \cite{Pfe84}.
 
 The fourth order Hermite integration scheme used in our code requires both
 the particle acceleration ${\bf a}_i$ and its time derivative ${\bf j}_i$
 (also called `jerk'). We integrate the evolution of the star cluster in a
 reference frame which is co-rotating with the bar, with a constant pattern
 speed $\Omega_{\mathrm {bar}}$.  The resulting equations of motion for a
 particle are then given by, e.g., \cite{BT84}:

\begin{equation}
 {\bmath a}_i = {\bmath a}_{\mathrm{cl}}
                - {\bmath \nabla} \Phi_{\mathrm{gal}}
                - 2 \left( {\bmath \Omega}_{\mathrm{bar}} \times {\dot{\bmath r}}_i \right)
                - {\bmath \Omega}_{\mathrm{bar}} \times \left( {\bmath \Omega}_{\mathrm{bar}} \times {\bmath r}_i \right) \ ,
\end{equation} 

 \noindent where $\bmath{a}_{\mathrm{cl}}$ is due to the self-gravity of the
 stellar cluster, $\Phi_{\mathrm{gal}}=\Phi_{\mathrm D} + \Phi_{\mathrm B} + \Phi_{\mathrm{bar}}$ and
 the last two terms are the Coriolis and centrifugal force, respectively. 
 The jerk is then given as:

\begin{equation}
 {\bmath j}_i = {\bmath j}_{\mathrm{cl}}
                - \frac{\partial}{\partial t} {\bmath \nabla}
                  \Phi_{\mathrm{gal}} 
                - 2 \left(  {\bmath \Omega}_{\mathrm{bar}} \times {\bmath a}_i \right)
                - {\bmath \Omega}_{\mathrm{bar}} \times 
                  \left( {\bmath \Omega}_{\mathrm{bar}} \times {\dot {\bmath r}}_i \right) \ .
\end{equation} 

 For the $N$-body realisation of the star clusters we use distribution functions
 following a Plummer density profile \citep{Plu11} or a King density profile
 \citep{King66}. We found no qualitative differences in the evolution between
 the two types of models in our set of simulations. Since our simulations
 with King models have been run with a higher particle number and also have been
 carried out to longer simulation times we limit ourselves in this publication
 to those models. To set up the (non-rotating) isotropic King model for a given
 core radius $r_{\mathrm{c}}$ and central concentration $W_{0}$ we use Walter
 Dehnen's {\tt mkking} routine from the open source {\sc Nemo} software package
 \citep{Teu95}.  

 All simulations presented in this work have been carried out on the dedicated {\sc Gpu}
 cluster {\sc Kolob} at the University of Heidelberg and at the {\sc Gpu} facilities at
 the Laboratoire d'Astrophysique de Marseille.

%--------------------------------------------------------------------------------------------
\subsection[]{Parameter and units}  
%--------------------------------------------------------------------------------------------
 
 We set the gravitational constant $G$ to unity and the units of mass and length to
 $M_{\mathrm u}=10^{6}\, {\mathrm{M}}_{\sun}$ and $R_{\mathrm u}=1\,{\mathrm{pc}}$,
 respectively. With this choice, the resulting units of time and velocity become 
 $\tau_{\mathrm u}=15 \times 10^{3}\,{\mathrm{yr}}$ and $v_{\mathrm u}=65.6\,{\mathrm{km s^{-1}}}$.
 Although the $N$-body models are intrinsically scale-free, one should bear in mind
 that both the galaxy and the star cluster must be scaled simultaneously when scaling
 to different units, e.g., physical units. Especially, a rescaling of the models will
 also change the bar pattern speed. Throughout this work we stay with the model
 units and refer to physical units only when necessary.

 Our choice of model parameter is based on the well studied galaxy model described 
 in \cite{SPA02a}. The corresponding mass distribution in this model results in
 a rotation curve as typically observed for disc galaxies. The parameter converted
 to our set of units are given in Tables~\ref{table:tab1} and \ref{table:tab2} for
 the galaxy and for the bar, respectively. The pattern speed of the bar is 
 correspondingly set to $\Omega_{\mathrm{bar}} = 7.64 \times 10^{-4} \mbox{ rad\,}
 \tau_{\mathrm u}^{-1}$ with the bar rotating counter-clockwise in the (non-rotating)
 lab-frame. The rotation period of the bar corresponds to
 $T_{\mathrm{bar}} \approx 8224 \,\tau_{\mathrm u}$.

\begin{table}
 \begin{minipage}{\columnwidth}
 \caption{Galaxy model parameter in model units.}
 \label{table:tab1}
 \begin{center}
 \begin{tabular}{lccc}
 \hline
 Component & $M_{0}\ [\times 10^4]$ & $A\ [\times 10^3]$ & $B\ [\times 10^3]$   \\
 \hline
 disc         &  16.4   &  3.0  & 1.0 \\
 bulge/halo   &   1.6   &  --   & 0.4 \\
 \hline
 \end{tabular}
 \end{center}
\medskip $M_{0}$ is the mass of the components given in the first column. $A$ and
 $B$ are the radial and vertical scaling parameters of the Miyamoto-Nagai potential.
\end{minipage}
\end{table}

\begin{table}
 \begin{minipage}{\columnwidth}
 \caption{Bar model parameters in model units.}
 \label{table:tab2}
 \begin{center}
 \begin{tabular}{lcccc}
 \hline
 Component & $M_\mathrm{bar} $ & $a_{\mathrm{bar}}$ & $b_{\mathrm{bar}}$  & $c_{\mathrm{bar}}$ \\
 \hline
 bar       &  $2 \times 10^4$ &  $6 \times 10^3$  & $1.5 \times 10^3$  & $0.6 \times 10^{3}$  \\
  \hline
\end{tabular}
 \end{center}
\medskip $M_{\mathrm{bar}}$ is the mass of the bar component. $a_{\mathrm{bar}}$, $b_{\mathrm{bar}}$
 and $c_{\mathrm{bar}}$ are the bar's 
 semi-principal axes.
\end{minipage}
\end{table}

 The parameters of the star cluster(s) are summarised in Table~\ref{table:tab3}.
 The half-mass radius of our (isolated) 50000 (50k) particle King model corresponds to
 $r_{\mathrm h} \approx 4$ and the tidal radius to about $r_{\mathrm t} \approx 34$.
 The total energy of the $N$-body system is about $E_{\mathrm{cl}}=-2 \times 10^{-3}$.

\begin{table}
 \begin{minipage}{\columnwidth}
 \caption{Star cluster parameters in model units.}
 \label{table:tab3}
  \begin{center}
 \begin{tabular}{cccccc}
 \hline
 $N$   & M$_{\mathrm{tot}}$ & W$_0$ & $r_{\mathrm c}$ & $\varepsilon$  \\
 \hline
 50000 & $0.2$  & 7.0   & 1.0   & 0.001 \\
 \hline
\end{tabular}
\end{center}
\medskip $N$ is the particle number used in the $N$-body model and M$_{\mathrm{tot}}$ is the total cluster mass.
 W$_0$ and r$_{\mathrm c}$ are the concentration parameter and the core radius of the King profile, respectively, 
 and $\varepsilon$ is the gravitational softening length used in the simulations.
\end{minipage}
\end{table}

%-------------------------------------------------------------------------------------------------------------------
\subsection[]{Planar periodic orbits}
%-------------------------------------------------------------------------------------------------------------------

\begin{figure}
\includegraphics[width=\columnwidth]{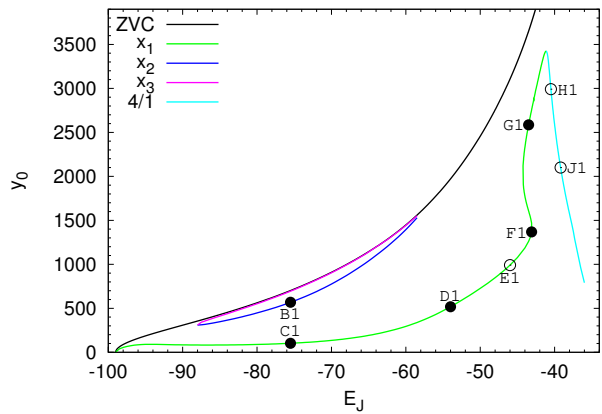}
\caption{Characteristic diagram of our barred potential. We plot the $y$-intercepts
 ($y_{0}$) of the main planar prograde periodic orbits as a function of Jacobi energy
 $E_{\mathrm{J}}$. The black line is the zero-velocity curve (ZVC). The main periodic 
 2-d families are shown in different colours as given in the legend. We mark the
 orbits on which we evolve the star clusters, indicating the stability by filled
 (stable) and open circles (unstable).
}
\label{fig:fig01}
\end{figure}

\begin{figure*}
 \begin{center}
  \includegraphics[width=0.45\textwidth]{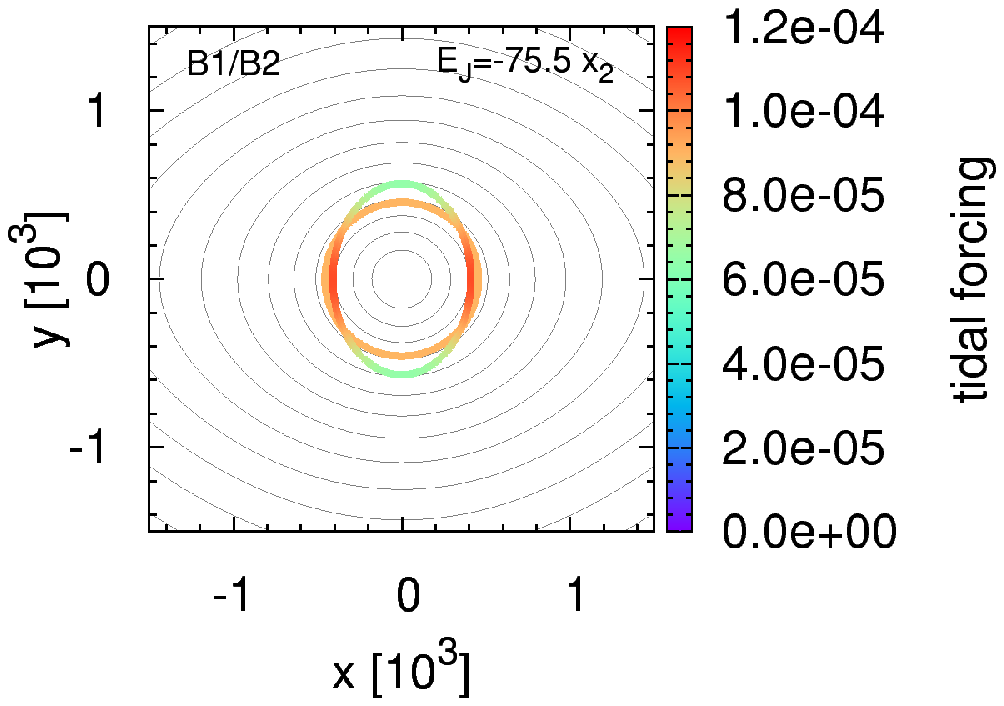}
  \includegraphics[width=0.45\textwidth]{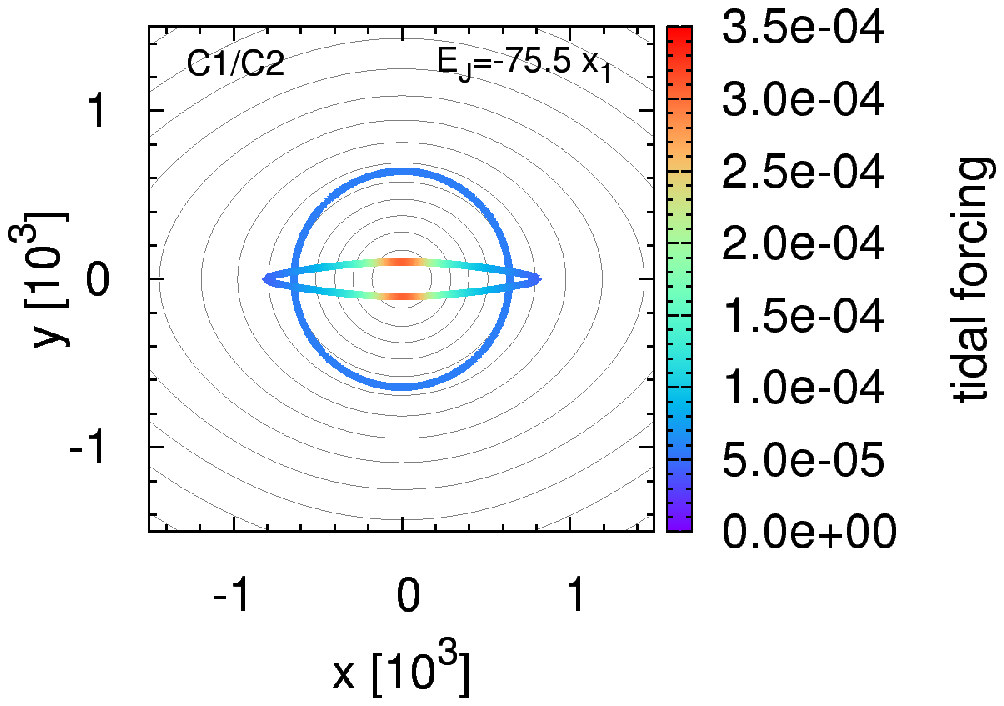} \\[-5ex]
  \includegraphics[width=0.45\textwidth]{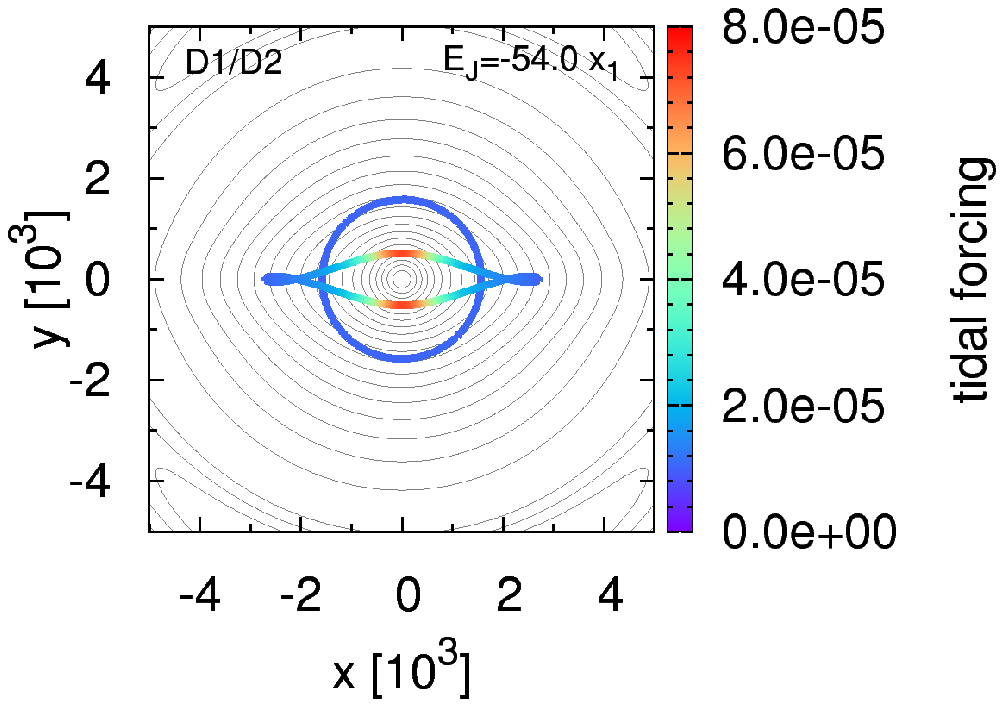}
  \includegraphics[width=0.45\textwidth]{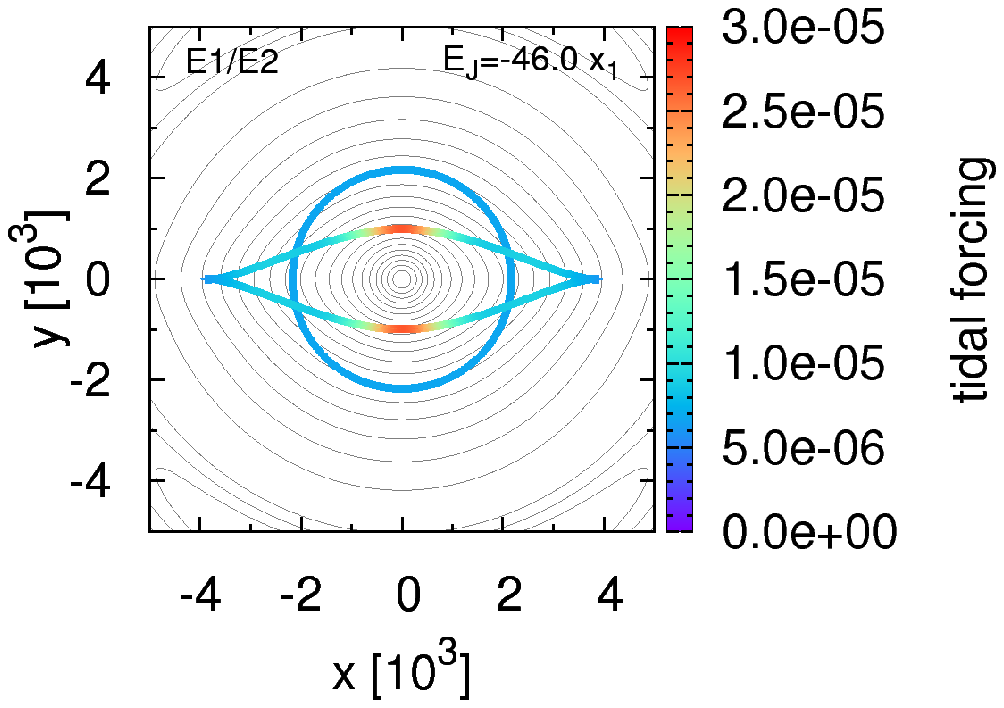} \\[-5ex]
  \includegraphics[width=0.45\textwidth]{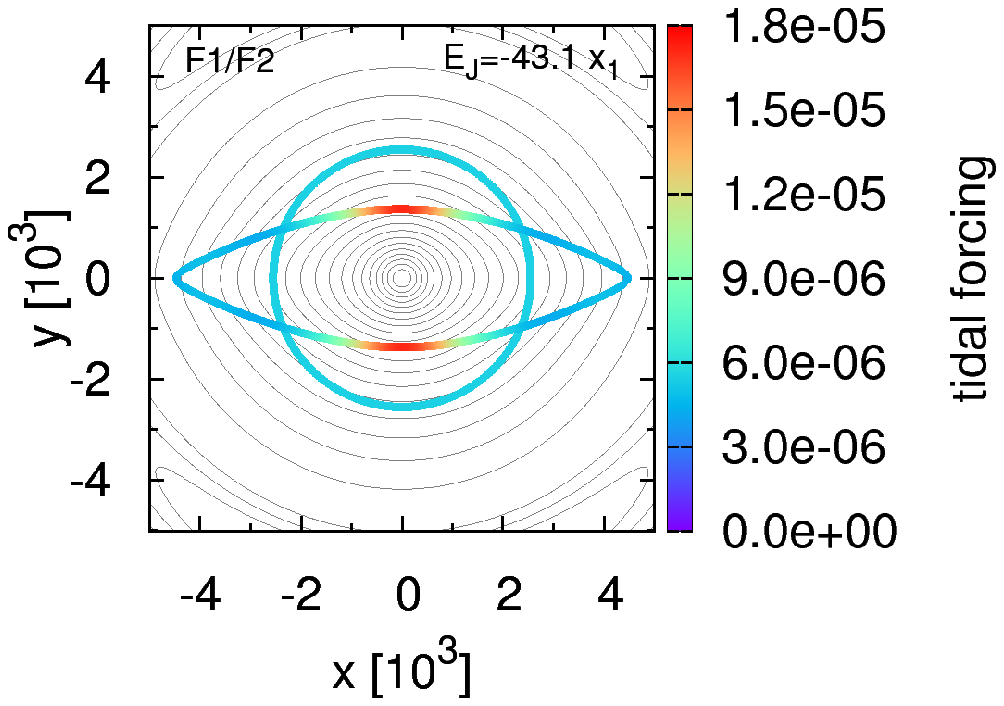}
  \includegraphics[width=0.45\textwidth]{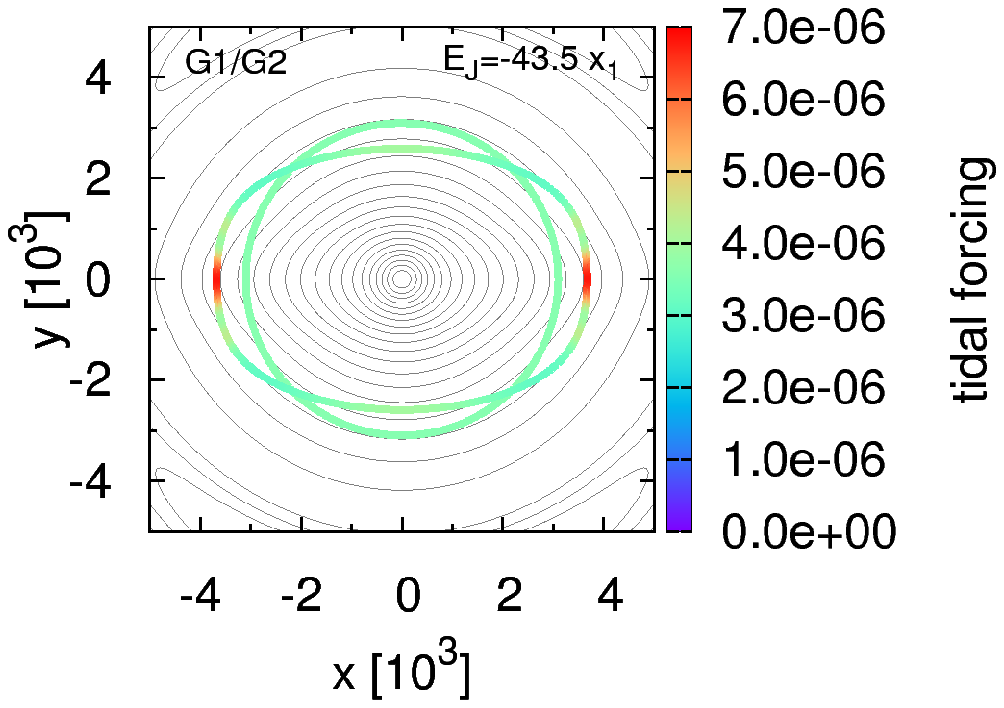} \\[-5ex]
  \includegraphics[width=0.45\textwidth]{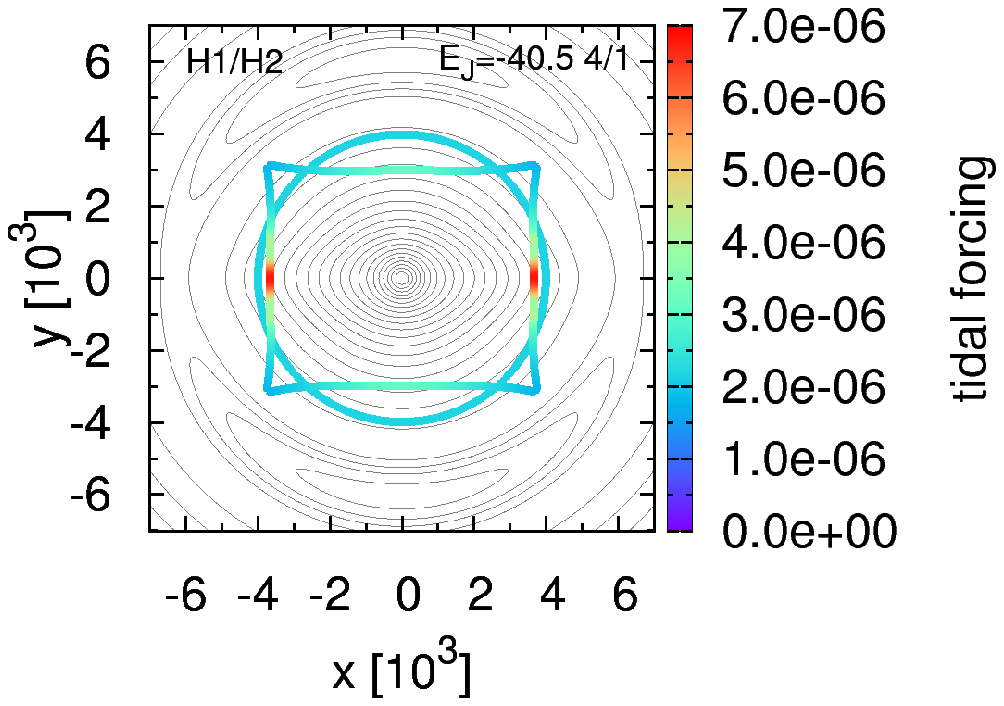}
  \includegraphics[width=0.45\textwidth]{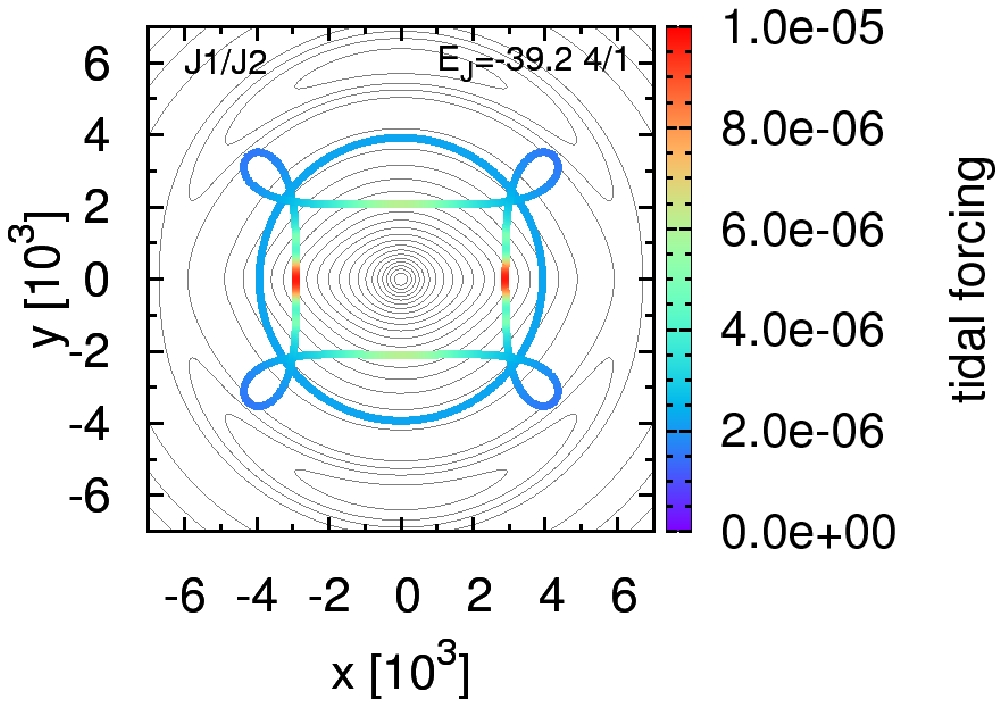}
 \end{center}
\caption{2-d planar periodic orbits in the barred
 and the axisymmetric potential. The corresponding orbit name, 
 Jacobi integral $E_{\rm J}$ (for the barred model) and family
 is given in each frame. The colour coding corresponds
 to the strength of the tidal forcing along the orbit. See main
 text (Section~\ref{sec:tidal}) for details. Iso-potential
 contours of the effective potential are plotted with grey lines.
}
\label{fig:fig02}
\end{figure*} 

 In this work we place the star cluster in the mid-plane of the galactic disc
 on an orbit which is periodic in the co-rotating reference frame of the bar. These orbits
 are the backbone of stellar bars (in 2-d) because they can confine regular regions
 (invariant tori) in phase-space around them. Orbits located in these regions show
 similar shapes and characteristics as their underlying parent periodic orbit. 
 Thus we would expect a similar evolution of the star cluster when placed on
 more general orbits.

 In Fig.~\ref{fig:fig01} we show the characteristic diagram for the main planar
 orbit families \citep[see, e.g.,][]{BT84} in the barred potential. We plot
 the $y$-axis intercepts ($y_{0}$, with $\dot{y}<0$) of the periodic orbits
 as a function of their Jacobi integral, $E_{\mathrm J}$, which is given as 
 $E_{\mathrm J} = E - J \,\Omega_{\mathrm{bar}}$. Here, $E$ and $J$ are,
 respectively, the specific energy and the angular momentum of the orbit.
 The Jacobi integral is a conserved quantity in the 
 frame corotating
 with the bar\footnote{Due to the self-gravity of the $N$-body system the Jacobi
 integral is not a conserved quantity in our simulations.}. 
 
 The green line shows the so-called $x_1$ orbit family, following the nomenclature
 of \citet{CP80}. These orbits are elongated along the bar major axis and close
 after one revolution in the bar potential and two radial oscillations (2/1). Close
 to the co-rotation radius the $x_1$ turns over to the four-periodic (4/1) orbit branch
 (cyan line). The blue and magenta line in Fig.~\ref{fig:fig01} represent the
 so-called $x_2$ and $x_3$ orbit families, respectively. These orbits are a class
 of 2/1 orbits which are elongated perpendicular to the bar major axis and
 are a signature of the presence of the inner Lindblad resonance(s).

 The set of orbits on which we place the star cluster in our simulations
 are marked and labelled in Fig.~\ref{fig:fig01}, where we indicate the 
 stability using different symbols. A detailed discussion of the orbits
 and their stability in the potential used here is given in \citet{SPA02a}.

 For comparison, we also run a second set of simulations for each of the selected
 orbits, putting the star cluster on a circular orbit about the galactic centre
 using an (nearly) axisymmetric potential. For the latter, we maintain the mass of
 the bar, but set the parameter $a$ in Eq.~\ref{eqn:bar} to $2.7 \times 10^{3}$,
 which is the mean value of the three axes as given in Table~\ref{table:tab2}.
 We then set $b=0.99\,a$ and $c=0.90\,a$ due to the required inequality of the
 three parameter (see Eq.~\ref{eqn:bar}). The axisymmetric galaxy components
 remain unchanged.

 The radius of the circular orbits is determined as follows: we calculate the
 orbit-averaged radial acceleration $\overline{a_{\mathrm R}}$
 on a test-particle in the
 barred potential\footnote{
   $\overline{a_{\mathrm R}} \equiv \frac{1}{N_j} \sum\limits_{j=1}^{N_j}
     {\bf a}_j \cdot {\bf e}_{{\mathrm R},j}$, where ${\bf a}_j$ is
     the test particles acceleration at time $t_j=(j-1) \Delta t$ and ${\bf e}_{{\mathrm R},j}$
     its normalised position vector. For a constant time-step $\Delta t$ one gets
     $N_j = T_\mathrm{orb} / {\Delta t} + 1$.
   }
 over one orbital period $T_{\mathrm{orb}}$ for each of the
 periodic orbits marked in Fig~\ref{fig:fig01}.
 Using a simple bi-section root-finding algorithm, we then determine the radius of
 the corresponding circular orbit which has the same mean radial force. 
 Information about both kinds of orbits (barred and circular) and about our set
 of simulations is given in Fig.~\ref{fig:fig02} and Table~\ref{table:tab4}.

 \begin{table}
 \begin{minipage}{\columnwidth}
 \caption{Simulations}
 \label{table:tab4}
  \begin{center}
 \begin{tabular}{llcrr}
 \hline
 Orbit   &   $E_{\mathrm J}$   &   family   &  $y_{0}$ $\left[10^3\right]$      &T$_{\mathrm {orb}}$   \\
 \hline
 A  & Isolation & --        &  --  & --      \\[0.2em]
 B1 &  -75.5    & $x_2$     & 0.57 &  812.3  \\
 B2 &  -75.2    & circular  & 0.46 &  707.5  \\[0.2em]
 C1 &  -75.5    & $x_1$     & 0.10 & 1041.4  \\
 C2 &  -68.5    & circular  & 0.64 &  946.0  \\[0.2em]
 D1 &  -54.0    & $x_1$     & 0.52 & 3178.8  \\
 D2 &  -49.1    & circular  & 1.57 & 2111.0  \\[0.2em]
 E1 &  -46.0    & $x_1$     & 0.99 & 4654.5  \\
 E2 &  -40.6    & circular  & 2.17 & 2773.9  \\[0.2em]
 F1 &  -43.1    & $x_1$     & 1.37 & 6075.4  \\
 F2 &  -36.6    & circular  & 2.55 & 3262.1  \\[0.2em]
 G1 &  -43.5    & $x_1$     & 2.59 & 8008.6  \\
 G2 &  -32.2    & circular  & 2.09 & 3985.0  \\[0.2em]
 H1 &  -40.5    & $4/1 $    & 2.99 & 14733.1 \\
 H2 &  -26.3    & circular  & 3.99 & 5167.0  \\[0.2em]
 J1 &  -39.2    & $4/1 $    & 2.10 & 14782.0 \\
 J2 &  -26.7    &  circular & 3.92 & 5083.5  \\
 \hline
\end{tabular}
 \end{center}
\medskip The first column gives the orbit name for each selected orbit.
 $E_{\mathrm J}$ is the Jacobi integral. The third column shows the
 corresponding orbit family. $y_{0}$ is the $y$-axis intercept of the
 orbit and T$_{\mathrm {orb}}$ is its orbital period.
\end{minipage}
\end{table}

%---------------------------------------------------------------------------------
\section{Results}
\label{sec:results}
%---------------------------------------------------------------------------------

 In this section we describe the evolution of the star clusters first in isolation
 and then when placed initially on a periodic orbit in our barred potential.
 We also compare the results with simulations of star clusters on circular orbits
 in the corresponding axisymmetric potential. Many of the features described in this
 section are better seen in animation, than in plots. For this reason we include
 animations of all the models described here under
 {\tt http://lam.oamp.fr/research/dynamique-des-galaxies/ scientific-results/star-cluster-evolution/}.

%-----------------------------------------------------------------------------------
\subsection{Simulation A - isolated star cluster} \label{sec:SimA}
%-----------------------------------------------------------------------------------

\begin{figure}
\begin{center}
\includegraphics[width=\columnwidth]{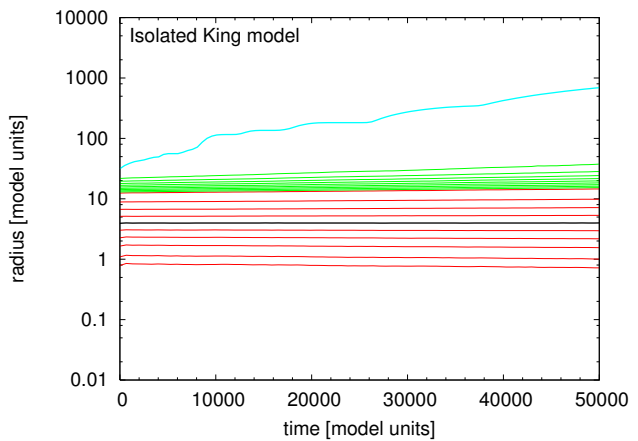}
\end{center}
\caption{Lagrange radii as a function of time for the isolated 50k King model
 (simulation A). Red solid lines show
 mass fractions of $5, 10, 20, 30, \ldots, 90$ per cent of the total cluster
 mass and green solid lines are fractions of $91, 92, 93, \ldots, 99$ per cent
 (from bottom to top). The half-mass radius is plotted with a black solid line.
 The cyan solid line corresponds to the {\em farthest} gravitationally bound
 cluster particle, i.e., the particle with the largest distance to the cluster centre
 and a negative total energy.}
\label{fig:fig03}
\end{figure}

 Before studying the star cluster evolution in the gravitational field
 of our three-component galaxy model, we first run a simulation of the
 isolated cluster, i.e., without any external gravitational force.
 
 In Fig.~\ref{fig:fig03} we show the Lagrange radii, i.e., spherical radii
 that contain a fixed fraction of mass, as a function of time for the
 isolated 50k King model (cf. Table~\ref{table:tab3}).
 We define the centre of the cluster as the centre of mass
 of all particles within a sphere of $r=0.2$ around the particle with
 the highest mass denisty. The latter has been calculated using routines
 from the software package {\sc{falcON}} \citep{Dehnen00, Dehnen02}.
 Due to close
 two- and few-body encounters between particles, its outer envelope
 slowly heats up and expands, while its inner regions simultaneously
 start to contract slowly. As can be seen from Fig.~\ref{fig:fig03},
 the model does not undergo core collapse during the integration
 interval of $50000$ time units, or some $750$\,Myr -- i.e., roughly
 $5$ bar rotations.

 We find that all stellar particles remain gravitationally bound to the
 isolated cluster until the end of the simulation. This is most likely
 due to the small gravitational softening which prevents particles to
 reach escape velocities during close stellar encounters. The introduction
 of a gravitational softening in our models slightly reduces the strength 
 of close particle encounters and prevents the formation of bound binaries.
 Since our simulations end clearly before core-collapse the dynamical
 impact of binaries can safely be neglected.

 For this isolated King model the total energy $E_{\mathrm{cl}}$ of the
 cluster is a conserved quantity and can be used to check the accuracy
 of the numerical integration. We calculate the relative energy error
 and find that the energy is conserved to better than $5 \times 10^{-3}$
 per cent over the full integration time.

%---------------------------------------------------------------------------------
\subsection{Periodic orbits - simulations B -- J}
%---------------------------------------------------------------------------------

 In this subsection we present a mainly qualitative description of our
 simulations. We will use these results to quantify in the next subsection
 certain aspects of the cluster evolution and the formation and evolution
 of the tidal structures.

%----------------------------------------------------------------------------------------------------------------
\subsubsection{Simulations C1 and C2}
\label{subsub:C1C2}
%----------------------------------------------------------------------------------------------------------------

\begin{figure}
\begin{center}
\includegraphics[width=\columnwidth]{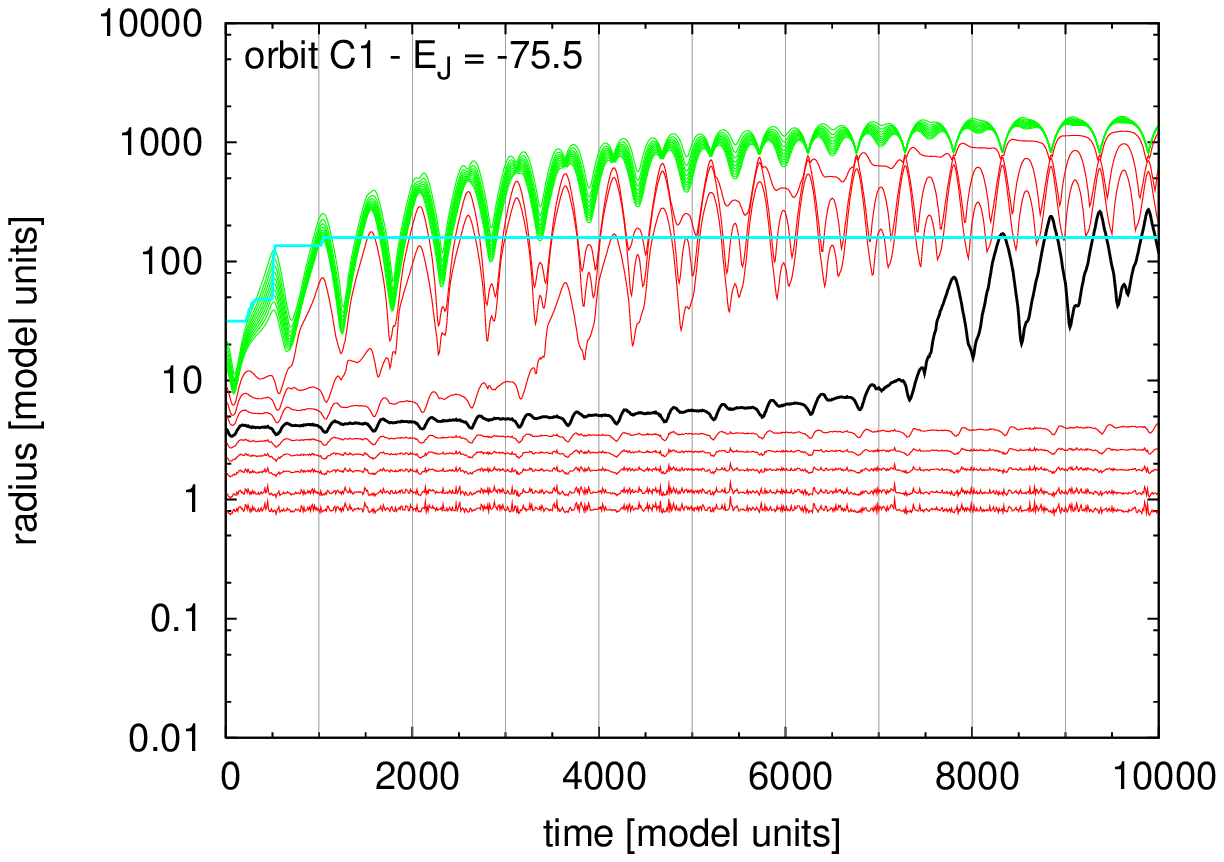}
\includegraphics[width=\columnwidth]{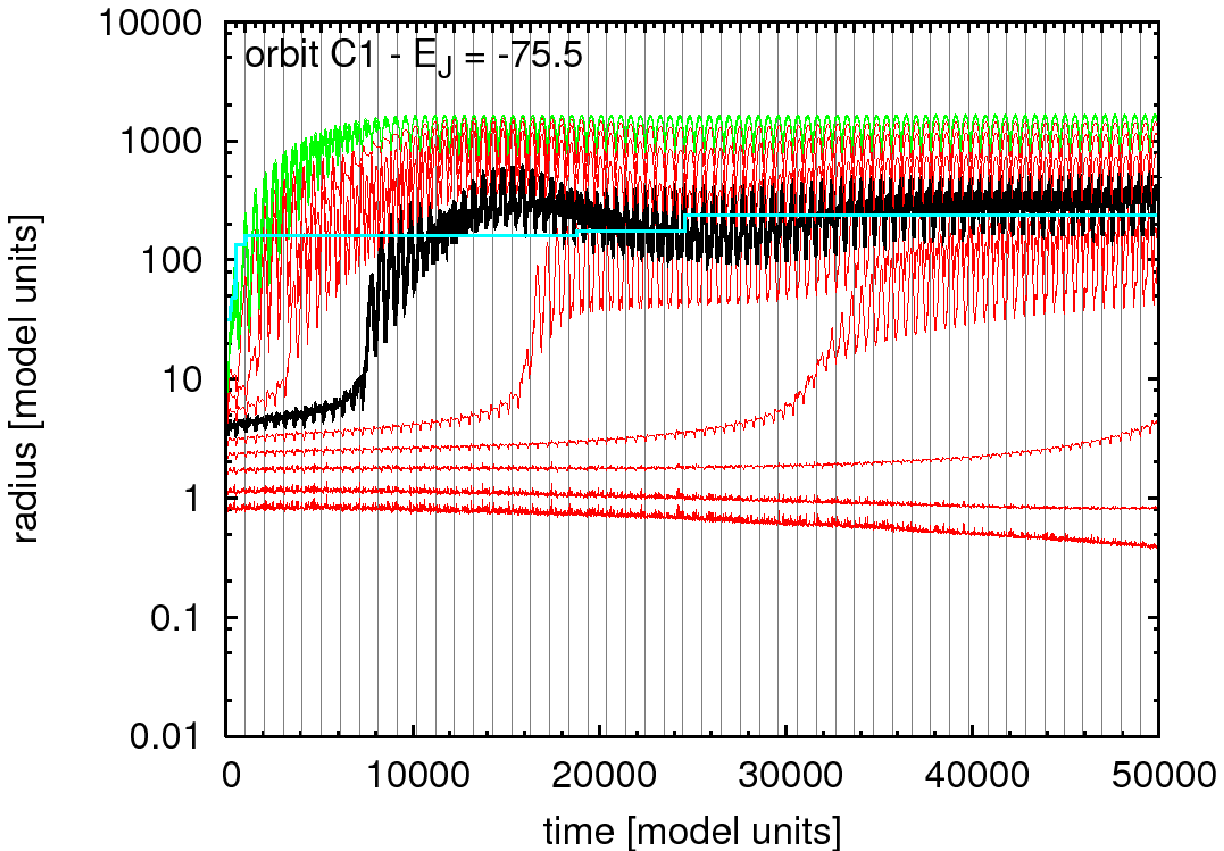}
\end{center}
\caption{Lagrange radii of the cluster in simulation C1 (cluster on an
 $x_1$ orbit with $E_{\mathrm J}=-75.5$) as a function of time. The grey
 vertical lines indicate the orbital period in the bar reference frame.
 The top and bottom panels show the time evolution up to $10000$ and
 $50000$ time units, respectively. The layout of both panels is the
 same as in Fig.~\ref{fig:fig03}.}
\label{fig:fig04}
\end{figure}

 The underlying orbit in simulation C1 belongs to the $x_1$ family
 (Figs.~\ref{fig:fig01} and \ref{fig:fig02}) and is dynamically stable.
 We place the cluster initially at pericentre at a distance from the
 galactic centre of about $R=y_0=100$ (Table~\ref{table:tab4}) in the
 mid-plane of the disc. The cluster is then launched from pericentre
 with an initial centre of mass velocity according to this $x_1$ orbit.
 Since it is started relatively close to the centre of the galaxy, it
 is initially not in dynamical equilibrium -- mainly due to the presence
 of strong tidal forces -- and adjusts itself in the very early phases
 of evolution. This effect is visible in Fig.~\ref{fig:fig04}, in which
 we show the evolution of the Lagrange radii as a function of time for
 this simulation.

\begin{figure}
\begin{center}
\includegraphics[width=\columnwidth]{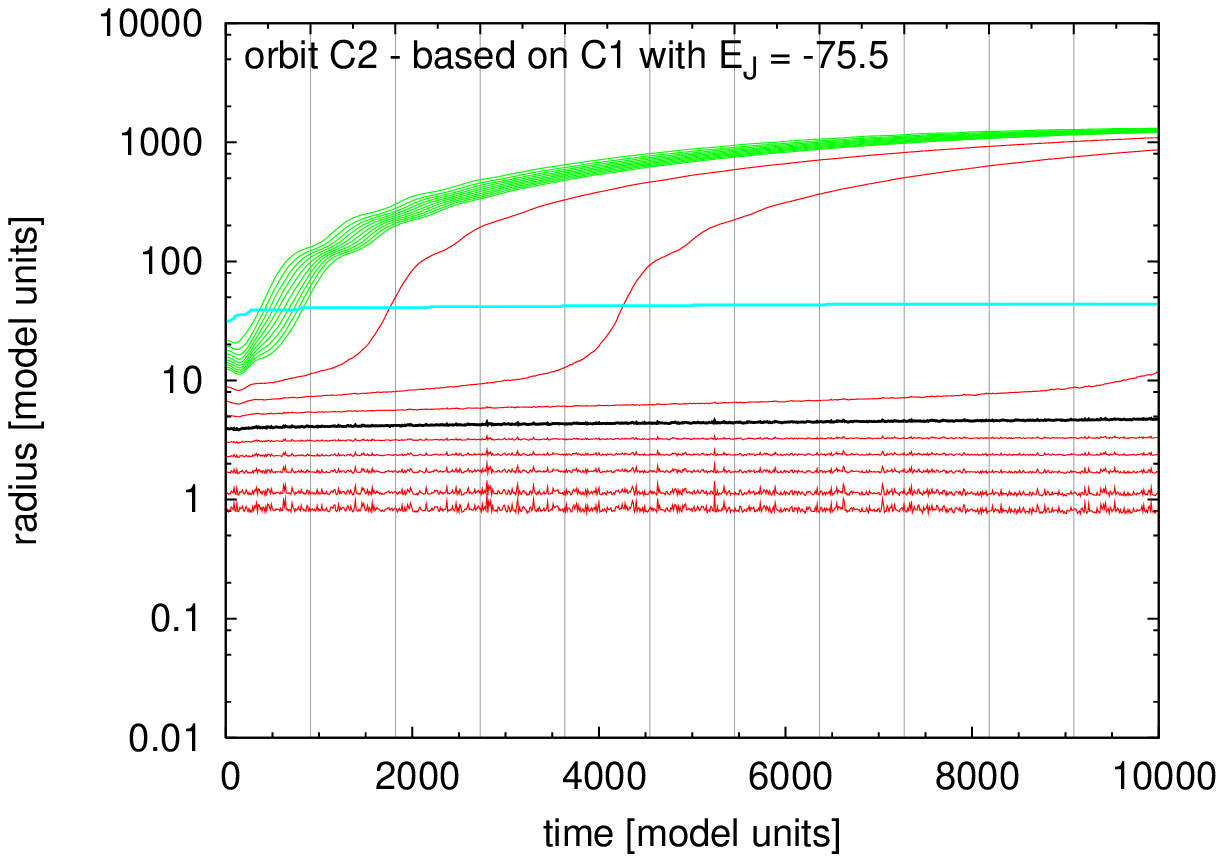}
\includegraphics[width=\columnwidth]{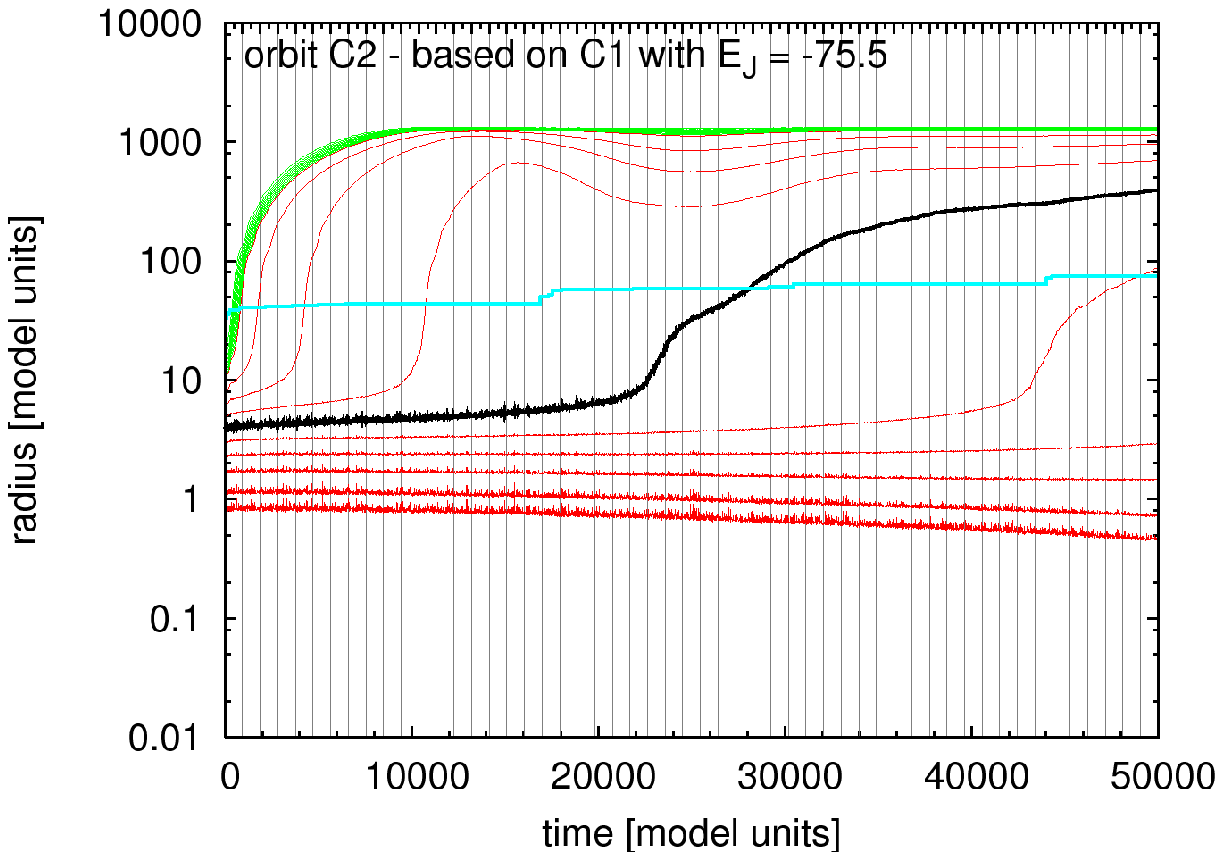}
\end{center} 
\caption{Lagrange radii for the simulation C2 (cluster on a circular orbit
 corresponding to orbit C1). Layout is the same as in Fig.~\ref{fig:fig04}.
 The sudden increase in the cyan line at around $t=18000$ results from a
 particle getting ejected from the cluster and thus is no more the
 last bound particle.} 
\label{fig:fig05}
\end{figure}

 After its first apocentre passage at $t \approx 260$ the cluster
 starts to form tidal tails which become clearly pronounced for
 the first time
 after about half an orbital period (see Table~\ref{table:tab4}).
 The leading and trailing tails are oriented along the underlying
 parent orbit and are initially confined to regions mainly inside
 and outside this orbit, separated by it. This is due to the fact
 that the particles are leaving the cluster -- depending on their
 Jacobi integral -- through the Lagrange
 points $L_1$ and $L_2$, where $L_1$ and $L_2$ lie between the cluster
 and the galactic centre and on the opposite
 cluster side, respectively \citep[see][for an illustration]{EJS09}.

 When the cluster approaches again the orbital apocentre, the stars
 forming the leading tidal tail start to slow down following their
 pericentre passage. This leads to a ``compression'' first of the leading
 tidal tail, and consecutively of the cluster itself and of the
 trailing tidal tail as they approach the orbital apocentre and
 slow down as well.
 The compression of the tidal tails is reflected in a decrease
 of the Lagrange radii (Fig.~\ref{fig:fig04}), which become minimal
 each time the cluster just passed the orbital apocentre. In this
 phase the trailing tail is still undergoing compression due to
 its deceleration, while the leading tail starts to expand again.
 The expansion of the tidal tails is reflected in an increase of
 the Lagrange radii which reach their maxima when the cluster passes
 pericentre.

 The effect of compression and expansion of the tidal features close
 to apocentre and pericentre passage has also been found in recent
 $N$-body simulations of star clusters moving on eccentric orbits
 in axisymmetric potentials \citep{KKBH10}. The compression/expansion
 of tidal streams and the resulting asymmetry when close to apocentre
 should thus be inherent to clusters moving on non-circular orbits. The
 oscillations found in the Lagrange radii appear with the same period
 as the one of the periodic orbit and are therefore linked to the
 variations of the tidal field along the trajectory of the cluster.
  
 As expected, the effect of periodic compression and expansion is indeed
 not found in our corresponding simulation C2  in which the star cluster
 moves on a circular orbit in the axisymmetric potential. In this case 
 the particles escape at an almost linear rate from the star cluster.
 For comparison with simulation C1 we show in Fig.~\ref{fig:fig05}
 the time evolution of the Lagrange radii for our model C2. The expansion
 of the different mass shells is relatively smooth
 and does not show any oscillations such as those found in model C1.
 This is due to the fact that the star cluster moves along the circular
 orbit with almost constant azimuthal velocity and is exposed to a constant
 tidal field. Note that the radius of the farthest unbound cluster particle 
 (see Fig.~\ref{fig:fig05}; cyan line) initially corresponds roughly to
 the initial tidal radius of the cluster $r_{\mathrm t}=35$, and slowly
 increases as the cluster is losing mass.

 The maximum value for the Lagrange radii is given in both simulations, C1 and
 C2, by the separation of the two apocentres. We find in our run C1 that the tidal
 tails fill about half the length of the $x_1$ orbit after  $t=4880$,
 or roughly $5$ orbital periods, and fills the full orbit at about $t=7800$,
 or about $8$ orbital periods.

 In both simulations, C1 and C2, sub-structures form 
 along the tidal tails in form of clumps (or {\em epicyclic over-densities})
 as described and analysed in previous publications
 \citep{KMH08,EJS09,KKBH10}. The position and strength of these clumps varies
 with the compression and expansion phases along the orbit. As the tidal
 tail reaches its maximum length at pericentre passage of the cluster, the
 relative clump density increases, i.e., shows the maximal density
 contrast between clumps and the tidal tails.

 The periodic density variations of the tidal tails are visible also
 in the oscillations of the outer Lagrange radii (e.g., those
 corresponding to mass fractions larger than $ 90$ per cent),
 with a period estimated by eye to about half the orbital period $T_{\mathrm{orb}}$
 (see Fig.~\ref{fig:fig04}). To show that the density
 variations are indeed correlated to the orbital period of the parent periodic
 orbit we use a Fast Fourier Transformation (FFT) to determine the oscillation
 period from the time series of the Lagrange radius $L_{98}$, i.e.,
 the radius which contains
 $98$ per cent of the initial cluster mass. For
 our run C1 we find an oscillation period to be $T_{\mathrm{L}}=529\pm1$,
 which in fact corresponds to roughly half the orbital period $0.5\, T_{\mathrm{orb}}$
 of the $x_1$ orbit. Our results for this analysis are summarised in Table~\ref{table:tab5}
 for all periodic orbits presented here.
 We find that at the end of the simulation (after $50000$ time units) the cluster
 in run C1 (barred potential) has lost about $79$ per cent of its initial mass,
 while the same cluster in run C2 (circular orbit) has lost about $62$ per cent
 of its initial mass. 
 
%-------------------------------------------------------------------------------------------------------------------
\subsubsection{Simulations B1 and B2}
%-------------------------------------------------------------------------------------------------------------------

\begin{figure*}
\begin{center}
\includegraphics[width=\columnwidth]{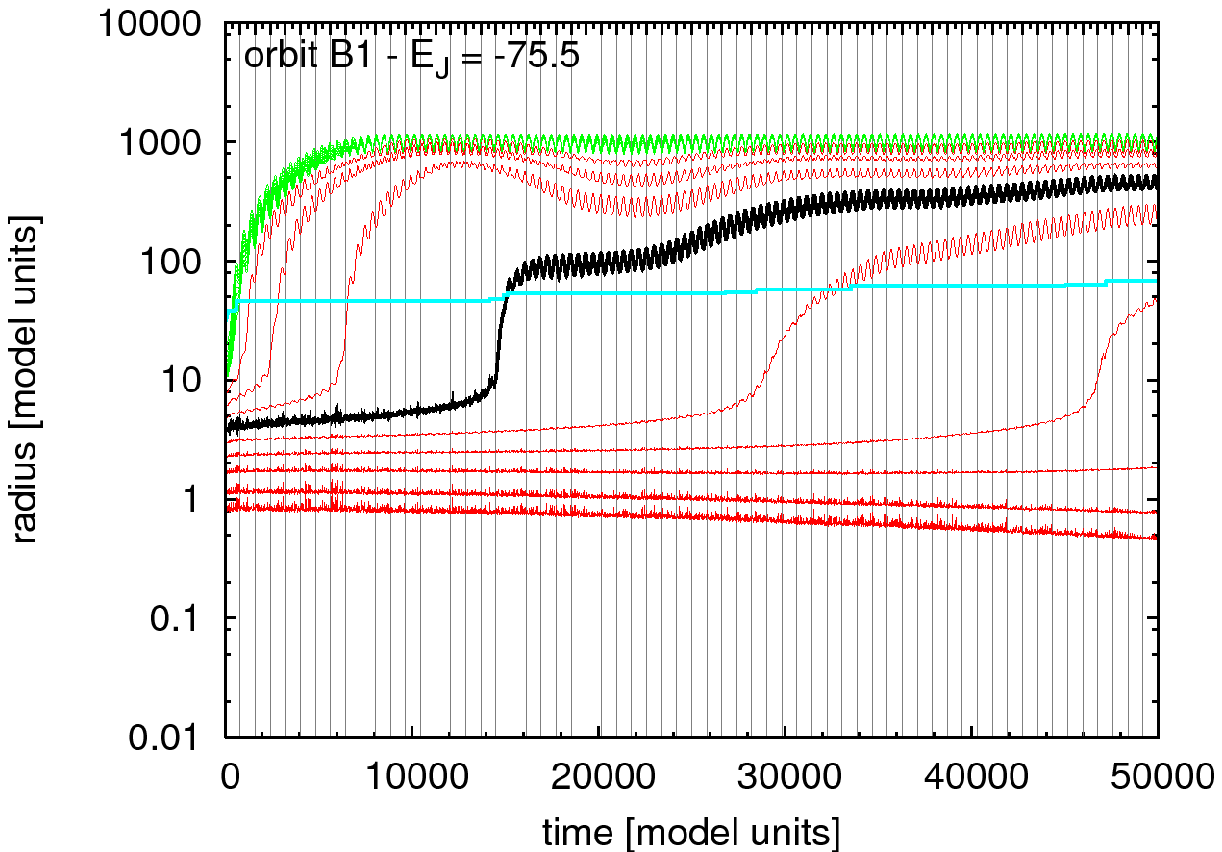}
\includegraphics[width=\columnwidth]{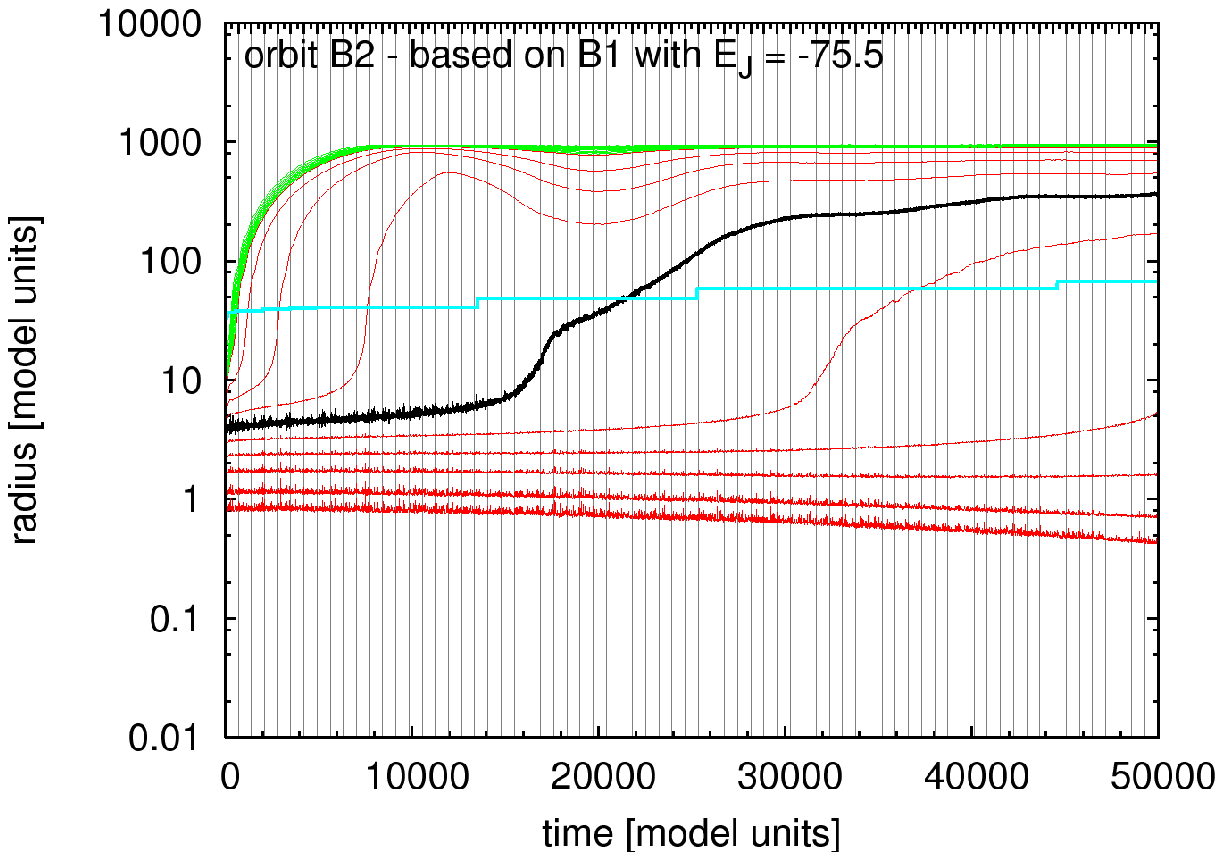}
\end{center}
\caption{Lagrange radii for the simulation B1 (left panel; cluster on
 an $x_2$ orbit with $E_{\mathrm J}=-75.5$) and simulation B2 (right panel;
 corresponding circular orbit).
 The layout of both panels is the same as in Fig.~\ref{fig:fig03}.}
\label{fig:fig06}
\end{figure*}

\begin{table}
 \begin{minipage}{\columnwidth}
 \caption{Periods.}
 \label{table:tab5}
 \begin{center}
 \begin{tabular}{clcrrc}
 \hline
 Orbit & $E_J$ & family & $T_{\mathrm{orb}}$  & $T_{\mathrm{L98}}$ & ratio  \\
 \hline
 C1 & -75.5    & $x_1$ &  1041.4 & $ 529\pm  1$ &  $\approx 2$ \\
 B1 & -75.5    & $x_2$ &   812.3 & $ 411\pm  1$ &  $\approx 2$ \\
 D1 & -54.0    & $x_1$ &  3178.8 & $1743\pm  9$ &  $\approx 2$ \\
 E1 & -46.0    & $x_1$ &  4654.5 & $2409\pm 18$ &  $\approx 2$ \\
 G1 & -43.5    & $x_1$ &  8008.6 & $4096\pm 51$ &  $\approx 2$ \\
 F1 & -43.1    & $x_1$ &  6075.4 & $3034\pm 28$ &  $\approx 2$ \\
 H1 & -40.5    & $4:1$ & 14733.1 & $3682\pm 41$ &  $\approx 4$ \\
 J1 & -39.2    & $4:1$ & 14782.0 & $3766\pm 43$ &  $\approx 4$ \\
 \hline
\end{tabular}
\end{center}
\medskip The six columns, from first to last, give the orbit name, its Jacobi
 integral, the corresponding orbit family, its orbital
 period T$_{\mathrm {orb}}$, the period of the expansion/compression
 measured using the Lagrange radius $L_{98}$ which contains 98 per cent of
 the initial cluster mass, respectively, and the ratio between the two
 periods.
\end{minipage}
\end{table}

 The parent periodic $x_1$ orbit in the previous simulation C1 is
 elongated along the major axis of the bar. We now describe the evolution
 of the cluster on an $x_2$ orbit with the same Jacobi
 integral of $E_{\mathrm J}=-75.5$, but which -- in contrast
 to $x_1$ orbits -- is oriented perpendicular to the bar
 major axis and is less eccentric. As before, this orbit
 is dynamically stable.

 Qualitatively, the evolution of the star cluster on the
 $x_2$ orbit (run B1) and on the corresponding circular
 orbit are found to be similar to those in simulations C1 and C2,
 respectively. As in the previous simulations we find formation
 of tidal tails and sub-structures therein in both kinds of models,
 i.e., in both the barred and the axisymmetric potentials.

 The effect of periodic compression/expansion of the tidal tails
 is less pronounced in run B1 (although clearly present) than in run C1
 since the underlying $x_2$ orbit is less
 eccentric. This is also reflected in the amplitude of the oscillations
 (compare Fig.~\ref{fig:fig06}; left panel). Again,
 the successive compression and expansion of the tidal
 tails correlates with the orbital period of the $x_2$
 orbit (see Table~\ref{table:tab5}).

 In simulation B2 (circular orbit) we note that the leading
 tidal tail and the trailing tidal tail are confined
 to regions inside and outside the underlying circular
 orbit owing to the particles epicyclic motion. Due to
 the strong non-linear perturbation of the bar in our
 simulation B1 the (linear) epicyclic approximation does
 not apply to the motion of the particles in the barred
 potential. As result of this we find a higher fraction
 of particles crossing the underlying periodic orbit as
 compared to the circular orbit B2.

 In simulation B1 the tidal tails fill one full orbit at
 about $t\approx 15000$. 
 The mass loss of the initial cluster mass is about
 $73$ percent in run B1 and about $69$ percent in
 run B2. In subsection~\ref{sec:tidal}, we will quantify and discuss
 this difference in terms of the tidal field for all our simulations.

%-------------------------------------------------------------------------------------------------------------------
\subsubsection{Simulations D1 and D2}
%-------------------------------------------------------------------------------------------------------------------

 In contrast to the previously described simulations, orbit D1 shows
 pronounced loops close to its apocentre (see Fig.~\ref{fig:fig02}),
 which is a clear signature for a strong bar \citep{Ath92}. The orbit
 is stable but lies in the close vicinity of a short instability
 strip in the $x_1$ characteristic curve (see Fig.\ref{fig:fig01}).

 We find that the formation and evolution of the tidal tails in simulations
 D1 and D2 is for several orbital periods qualitatively similar to that
 of cases with lower Jacobi integrals.
 Initially the forming tidal tails follow closely the underlying periodic
 orbit. After several apocentre passages of the cluster, however, we find
 that the tidal tails become a bit more diffuse compared to the previously
 described simulations. We attribute this effect to the dynamical
 stability of the 
 periodic orbit. In the particular case of simulation D1 the underlying periodic
 orbit is dynamically stable. Particles in the tidal tails which leave the
 cluster, however, enter regions in phase space which correspond to an
 unstable periodic orbit and are
 thus not confined to the invariant tori of the initial parent orbit anymore.

\begin{figure}
\begin{center}
\includegraphics[width=\columnwidth]{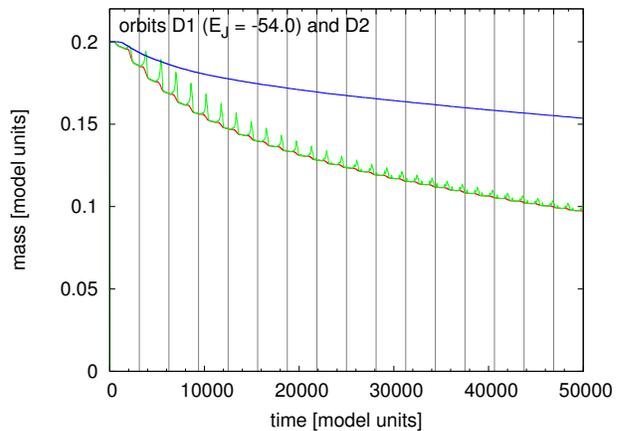}
\end{center}
\caption{Mass loss from clusters in simulations D1 and D2. Red and blue line: 
 Remaining cluster mass in D1 and D2, respectively, after removing
 all particles which have ever crossed the sphere around the
 cluster density centre with radius $2 \times R_{\mathrm{cut}}$. Green line:
 Mass of all particles in D1 that reside within this sphere at the current time.}
\label{fig:fig07}
\end{figure}

\begin{figure*}
\begin{center}
\includegraphics[width=\columnwidth]{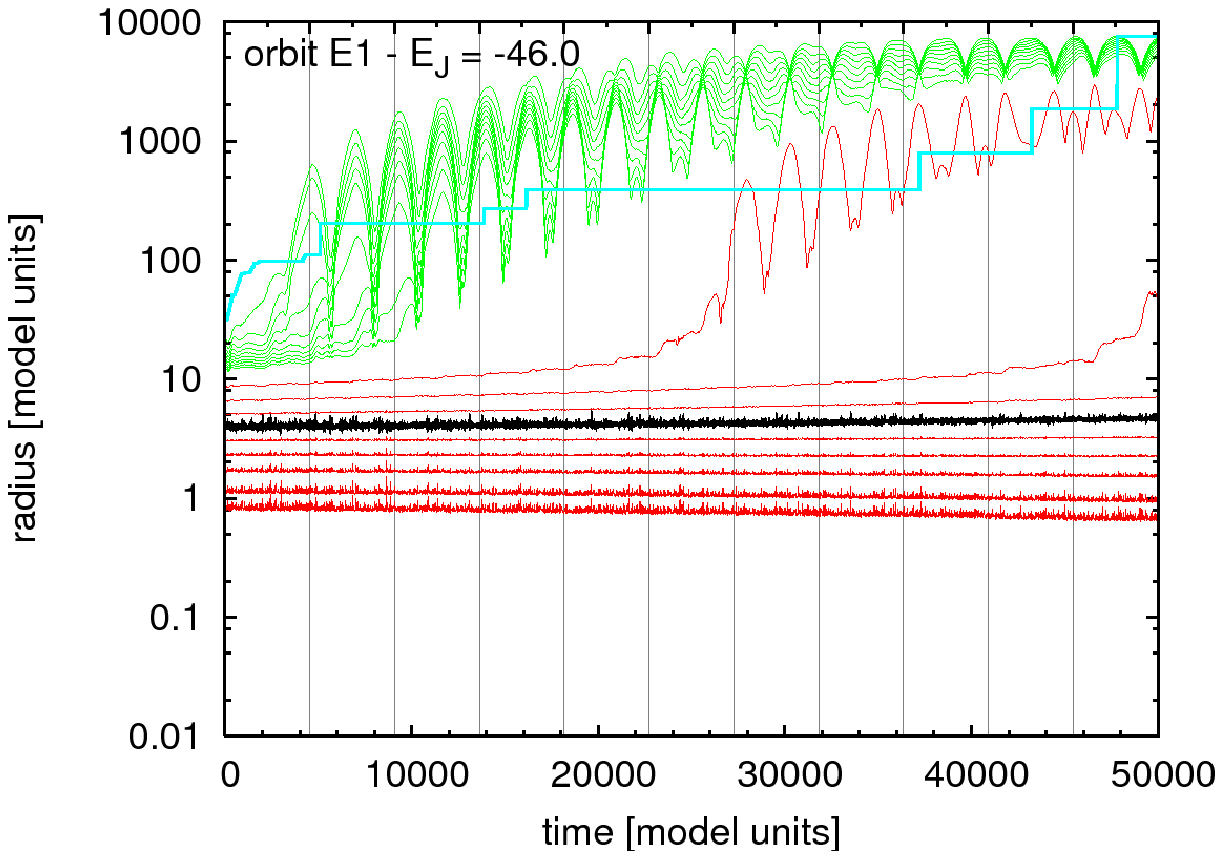}
\includegraphics[width=\columnwidth]{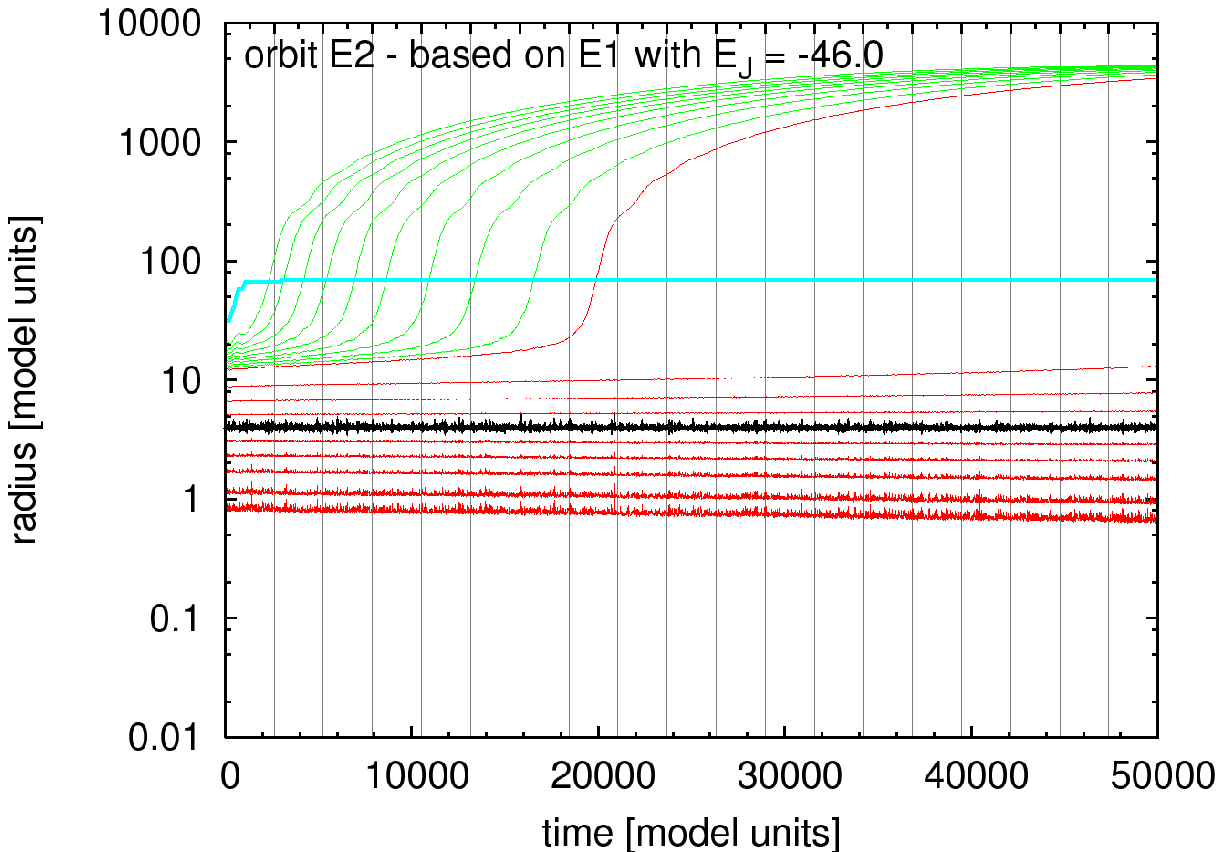}
\end{center}
\caption{Lagrange radii for the simulation E1 (left panel; cluster on
 an $x_1$ orbit with $E_{\mathrm J}=-46.0$) and simulation E2 (right panel,
 cluster on the corresponding circular orbit).
 The layout of both panels is the same as in Fig.~\ref{fig:fig03}.}
\label{fig:fig08}
\end{figure*}

 As before, we find oscillations in the Lagrange radii (not shown here) whose
 period correlates with the orbital period  of the D1 orbit (see Table~\ref{table:tab5}).
 In Fig.~\ref{fig:fig07} we show the cluster mass in run D1 as a function of
 time. Here we use two definitions of the cluster members, namely, (a) particles
 which have never crossed a radius corresponding to twice the initial cut-off
 radius $R_{\mathrm{cut}}$ at any given time (Fig.~\ref{fig:fig07}:
 red line), and (b) all particles that reside within $2 \times R_{\mathrm{cut}}$
 at any give time (Fig.~\ref{fig:fig07}: green line). The oscillations
 found in the latter case have approximately the same period as that 
 derived from the Lagrange radius $L_{98}$. A similar behaviour 
 is also found (not shown here) in the previously described runs
 B1 and C1. We note that the mass loss $f_{\mathrm{ml}}(t)$ in all our
 simulations consists of two main contributions, i.e., an initial
 exponential decay $f_1(t)$ continuing with a secular linear decay $f_2(t)$:

\begin{equation} \label{eq:massloss}
 f_{\mathrm{ml}}(t) = f_1(t) + f_2(t) = (m_0 \, e^{-m_1 \, t}) + (m_2 - m_3 \, t)  \, .
\end{equation}

\noindent We discuss this dependency in more detail in Sec.~\ref{sec:tidal}. At
 the end of simulation D1 about $52$ percent of the initial cluster mass is
 gravitational unbound (see Fig.~\ref{fig:fig07}), while for the corresponding
 circular orbit in simulation D2 about $23$ per cent of the original cluster
 mass is unbound.

%-------------------------------------------------------------------------------------------------------------------
\subsubsection{Simulations E1 and E2}
%-------------------------------------------------------------------------------------------------------------------

\begin{figure}
\begin{center}
\includegraphics[width=\columnwidth]{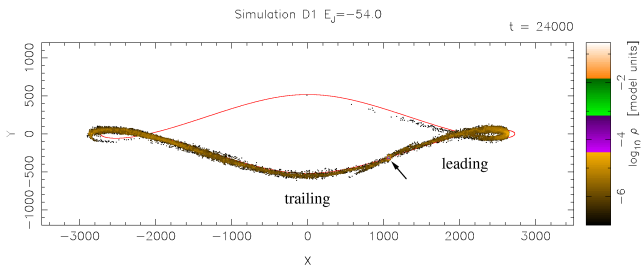}
\end{center}
\caption{Star cluster in simulation D1 at time $t=24000$. The particles are
 coloured according to the local mass density of the cluster. The underlying
 periodic is shown with a red line. The arrow points to the cluster
 position. Leading and trailing arms are labelled accordingly.
 }
\label{fig:fig09}
\end{figure}

\begin{figure}
\begin{center}
\includegraphics[width=\columnwidth]{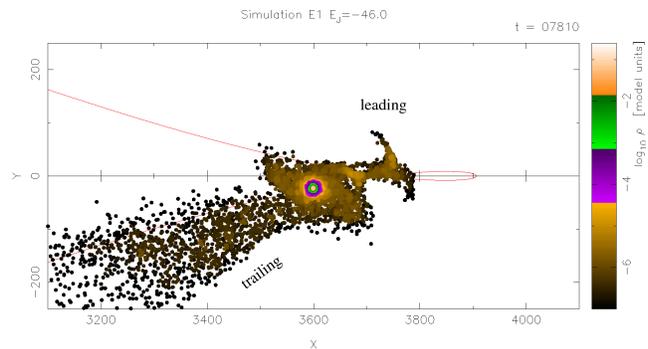}
\end{center}
\caption{Star cluster in simulation E1 at time $t=7810$. The particles are
 coloured according to the local mass density of the cluster. The underlying
 periodic orbit is shown with a red line.}
\label{fig:fig10}
\end{figure}

 To test our hypothesis of the influence of the stability of the
 periodic orbits on washing out the tidal tails we chose now 
 an unstable $x_1$ orbit roughly centred on the instability strip 
 in the characteristic curve between $E_{\mathrm J} \approx -48.4$ and $-43.6$.
 As in the previous case, this $x_1$ orbit shows loops close to its
 apocentres (see Fig.~\ref{fig:fig02}). In Fig.~\ref{fig:fig08}
 we show the evolution of the Lagrange radii for simulations E1 and E2.
 The initial evolution
 of the cluster and its tidal tails follow the underlying parent orbit.
 After a few apocentre passages, however, the structure of the trailing
 tidal tail becomes more diffusive and even shows multiple tail-like
 structures (e.g., compare Figs.~\ref{fig:fig09} and ~\ref{fig:fig10}).
 Both the leading and the trailing tidal tails wash out after 
 a few apocentre passages, while the cluster on the corresponding
 run E2  (circular orbit) shows nicely separated tidal tails.

 The clusters in E1 and E2 have lost by the end of the simulation about $30$ and
 $17$ per cent of the initial cluster mass, respectively. This supports the 
 trend that a star cluster in barred potential loses more mass than the
 same cluster on the corresponding circular orbit in the axisymmetric case, when
 being evolved for the same time period. 

%-------------------------------------------------------------------------------------------------------------------
\subsubsection{Simulations F1 and F2}
%-------------------------------------------------------------------------------------------------------------------

 The $x_1$ orbit in our simulation F1 has peak-like edges at apocentre.
 The orbit lies right at the knee (kink) of the $x_1$ characteristic curve
 in a stable region, but just next to an unstable region (towards
 larger pericentres). 

 The formation of the tidal tails and their evolution is similar to 
 that of the previous models. At $t \approx 7300$ the leading tidal tail is
 highly compressed into one leading clump with distance of
 $\Delta R \approx 150$. The star cluster in run F1 loses about
 $11$ per cent and in run F2 about $14$ per cent of its initial mass.

\begin{figure*}
\begin{center}
\includegraphics[width=\columnwidth]{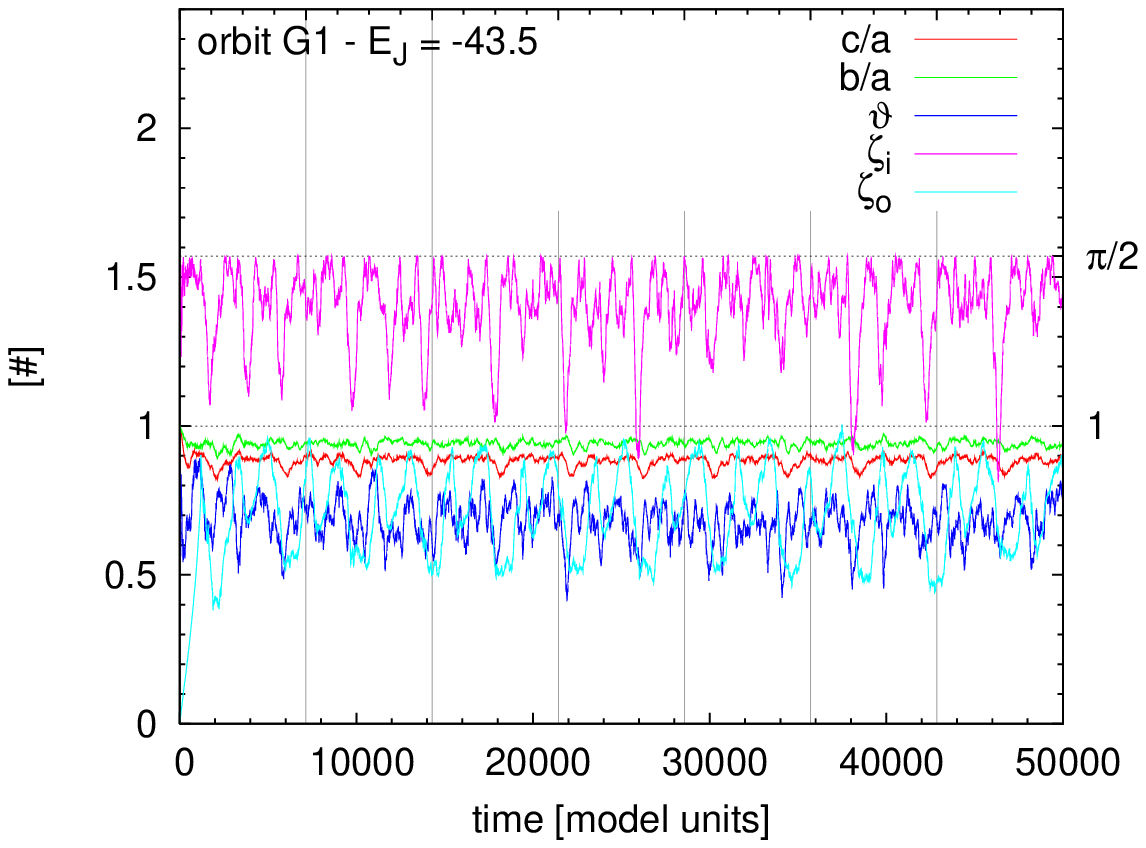}
\includegraphics[width=\columnwidth]{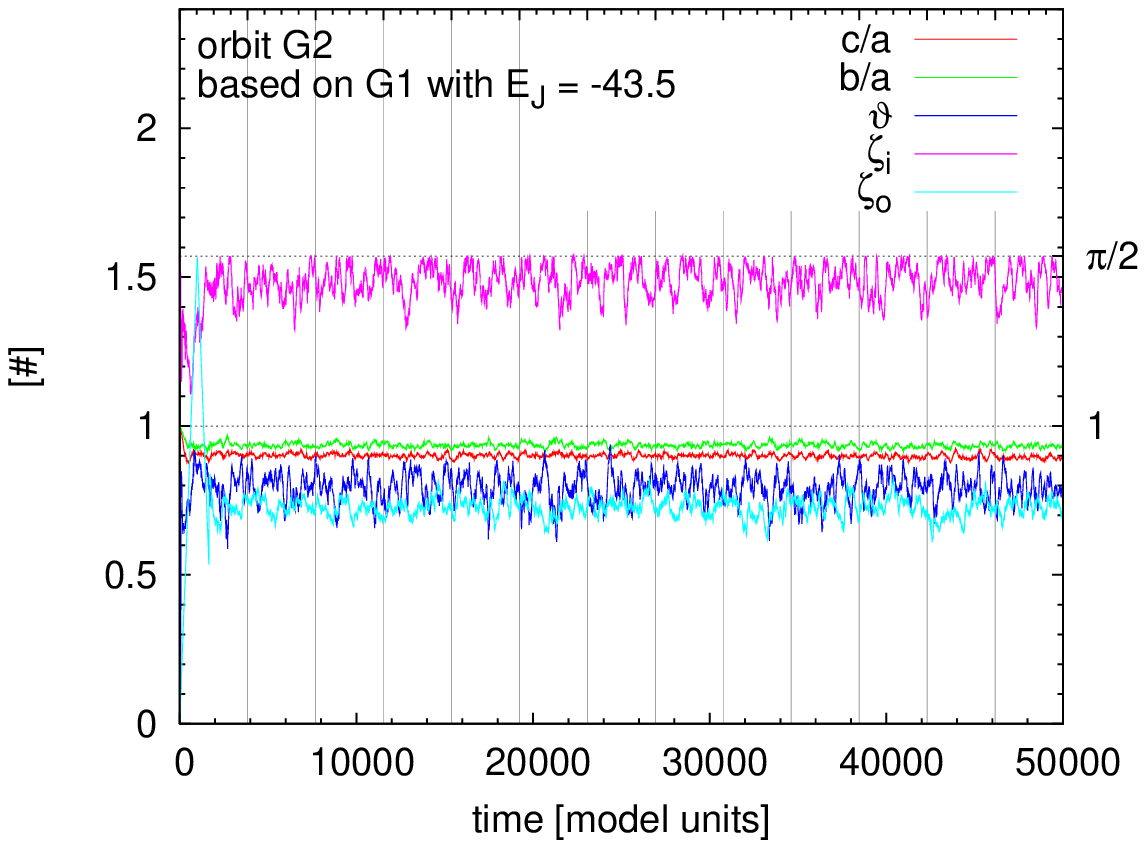}
\end{center}
\caption{Axial ratios $c/a$ (red line) and $b/a$ (green line) derived from moments of
 inertia within a spherical shell with $r=5\ldots25$, as well as the corresponding triaxiality
 parameter $\vartheta$ (blue line) as given in Eq.~\ref{eqn:triax}. 
 We also show the angle between the clusters centre of mass velocity vector and
 the major axis unit vector based on the moments of inertia measured (a) in shells
 with $r=5\ldots25$ ($\zeta_i$; magenta line) and (b) with $r=100-150$
 ($\zeta_o$; cyan line). Left panel: results from run G1. Right panel:
 results from run G2.}
\label{fig:fig11}
\end{figure*}

%-------------------------------------------------------------------------------------------------------------------
\subsubsection{Simulations G1 and G2}
%-------------------------------------------------------------------------------------------------------------------
 
 This orbit is a stable $x_1$ orbit. The formation of the tidal tails
 is similar to the case of circular orbits. The leading tail
 slows down when approaching apocentre and the clumpy sub-structure is
 stronger in this phase. The location of the clumps within the
 tails changes with time. Indeed, as the clumps in the leading tail
 slow down when approaching 
 apocentre their relative distance with respect to the cluster centre becomes
 shorter.
 At the end of the simulation the tidal tails extend roughly over the underlying
 periodic orbit. As in the previous runs the centre of mass of the
 cluster does not stay on the underlying orbit. The star cluster in
 simulations G1 and G2 loses about $19$  and $9$ per cent, respectively,
 of the initial cluster mass.

%-------------------------------------------------------------------------------------------------------------------
\subsubsection{Simulations H1 and H2}
%-------------------------------------------------------------------------------------------------------------------

 The orbit in simulation H1 is a $4/1$ orbit, i.e., a four-periodic
 orbit closing after one rotation and four radial oscillations.
 It is an unstable orbit and has sharp pointy edges at its apocentres.
 As before, we witness the formation of tidal tails, which after apocentre
 passages do not follow the underlying orbit anymore. The evolution 
 of the star cluster in this run further supports our hypothesis that the
 orbit stability in the global galactic potential determines the
 orientation and shape of the tidal tails and also how diffusive they
 are. The cluster in run H2, i.e., on the corresponding circular
 orbit, does not show any peculiarities as compared 
 to the earlier simulations using circular orbits. The mass loss at the
 end of the simulation is about $8$ per cent in run H1 and about $4$ per
 cent in run H2.

%-------------------------------------------------------------------------------------------------------------------
\subsubsection{Simulations J1 and J2}
%-------------------------------------------------------------------------------------------------------------------
 
 This orbit in simulation J1 is again an unstable $4/1$ orbit. In contrast
 to the previous orbit H1, however, it has loops at its apocentres. 
 The tidal tails which form in this model are initially aligned with
 the underlying periodic orbit. This changes after the first apocentre
 passage, when reaching the outer edge of the loop there.
 The tail rapidly swings around when the cluster runs through the loop
 and aligns again with the orbit after the cluster leaves the loop.
 After the third apocentre passage the tidal tails are
 less sharply defined and less aligned with the periodic orbit. As in the
 other cases with unstable orbits, the tidal tails slowly dissolve
 into a very diffusive morphology. We furthermore note that the 
 cluster itself does not remain centred on the initial periodic
 orbit, but its density centre reaches separations up to
 $\Delta R \approx 100$ from the initial underlying periodic orbit.

 Both the mass loss rate of the cluster and the density in the tidal
 tails is relatively low. At the end of the run the tidal tails
 have filled about half the underlying orbit. In simulation J1 
 the star cluster loses about $11$ per cent of its initial mass
 and about $5$ per cent in run J2.

%------------------------------------------------
\subsection{Shape and orientation of the cluster}
%------------------------------------------------

 In order to quantify the shape and orientation of both the cluster and
 the tidal tails formed in the simulations, we calculate the moments of inertia
 tensor of the cluster inside a spherical shell within $5 \geq r \geq 25$
 and determine the corresponding set of eigenvalues $I_a, I_b, I_c$,
 with $I_a \leq I_b \leq I_c$. For the analysis we use the {\sc Eispack}
 software package \citep{SBD76}.  From the eigenvalues we then determine
 the axial ratios as $b/a = \sqrt{(I_c+I_a-I_b)/(I_c+I_b-I_a)}$ and
 $c/a = \sqrt{(I_b+I_a-I_c)/(I_c+I_b-I_a)}$
 as well as the triaxiality parameter as defined, e.g., in \citet{FIZ91}, as

\begin{equation}
 \vartheta  = (1 - b^{2}/a^{2})/(1 - c^{2}/a^{2}) \ ,
 \label{eqn:triax}
\end{equation}

 \noindent  with $a > b > c$. In terms of the eigenvalues, Eq.~\ref{eqn:triax} can also
 be rewritten as $\vartheta = (I_b - I_a)/(I_c - I_a)$. This definition gives
 $\vartheta=1$ for a prolate object ($a > b = c$; cigar-like shape) and $\vartheta=0$
 for an oblate object ($a = b > c $; disc-like shape). An example of the time
 evolution of $\vartheta$ is shown in Fig.~\ref{fig:fig11} (left panel; blue line).
 We find that star clusters on orbits in barred potentials and high energies
 $E_{\mathrm{J}}$ typically
 have smaller triaxiality parameter $\vartheta$, while clusters on orbits with
 smaller energy $E_{\mathrm J}$ have values of $\vartheta \approx 1$.

 Furthermore, we determine the angles $\zeta_i$ between the current
 centre of mass velocity vector of the cluster and the major axis inertia eigenvector
 determined from the mass distribution of the cluster within a spherical shell
 between $r$ = 5 and 25. Similarly we define $\zeta_o$ using the
 corresponding eigenvector for the tidal tails (measured between
 $r$ = 100 and 150). Fig.~\ref{fig:fig11} (right panel) shows the
 results for the circular orbit G2.
 Once the tidal tails have formed, the axial
 ratios of the cluster remain roughly constant. From the moments of
 inertia we find that the cluster is elongated perpendicular
 (i.e., $\zeta_i \approx 0.5 \pi$)
 to  its trajectory (due to the radial tidal force). The tidal tails
 follow the underlying orbit. Within the region $r$=100 and 150 (which
 is relatively close to the cluster compared to the total length
 of the tails) we get $\zeta_o \approx 0.25 \pi$ for our run G2.

\begin{figure}
\begin{center}
\includegraphics[width=\columnwidth]{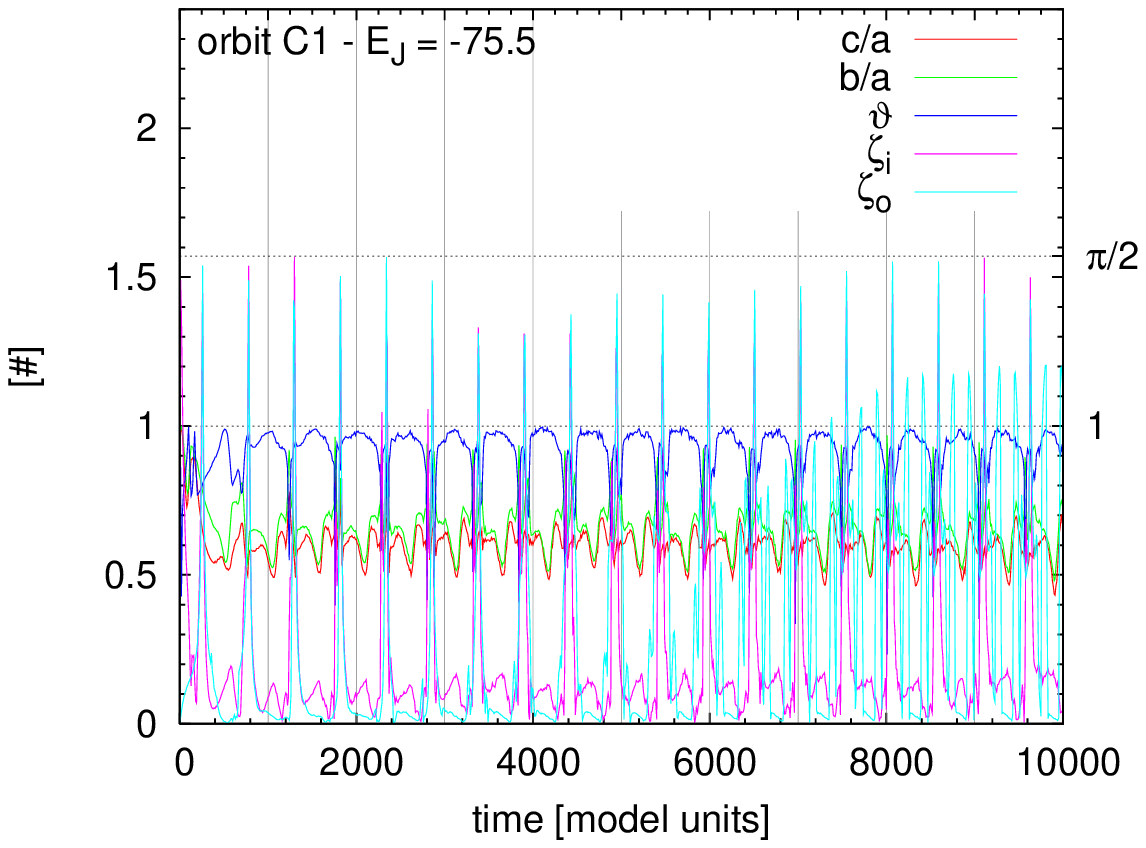}
\includegraphics[width=\columnwidth]{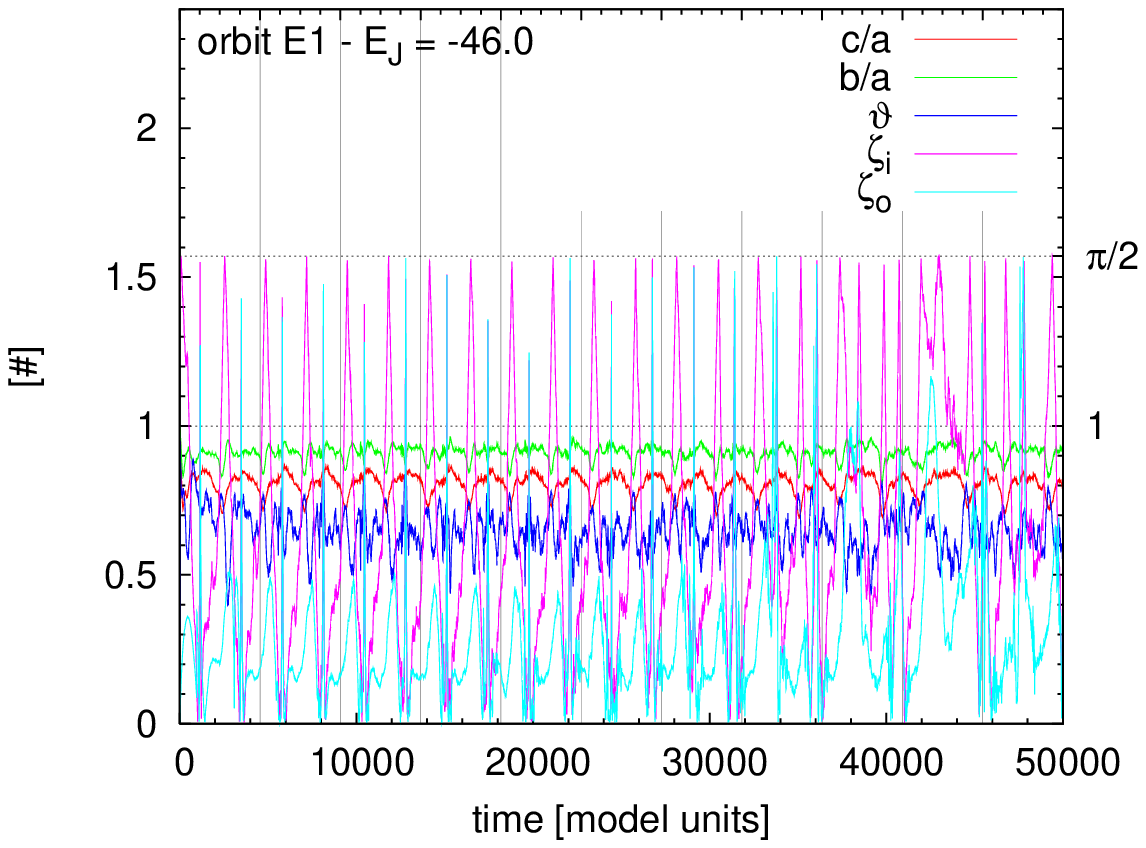}
\end{center}
\caption{Results from our simulations C1 (top panel) and E1 (bottom panel).
 The layout is the same as in Fig.~\ref{fig:fig11}.}
\label{fig:fig12}
\end{figure}
 
 For orbits in barred potential (see Fig.~\ref{fig:fig11}, left panel)
 the semi-major axis of the cluster lies in the orbital plane but periodically
 changes its orientation within the orbital plane with respect to the cluster
 trajectory. The same is true for the tidal tails. The oscillations
 of $\zeta_o$ have a larger amplitude than those of $\zeta_i$. 
 The evolution of the orientation of both the cluster and tidal
 tails is more regular as compared to orbits in the barred potential.
 We note that the amplitude of the oscillations in $\zeta_o$ are
 higher for orbits which have large variations in their radius
 of curvature (compare Fig.~\ref{fig:fig12}).

\begin{figure}
\begin{center}
\includegraphics[width=\columnwidth]{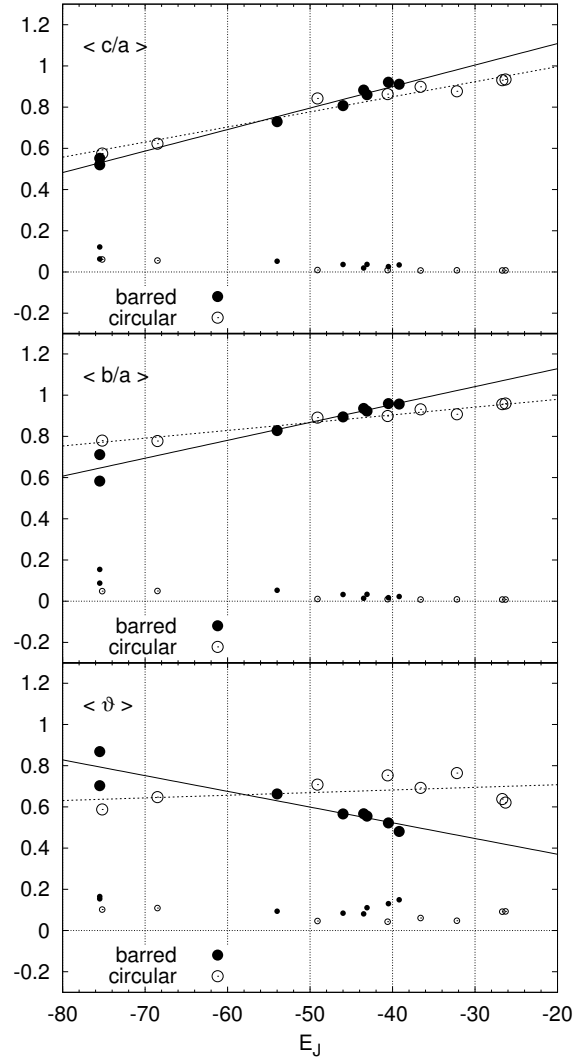}
\end{center}
\caption{Time-averaged axial ratios $<c/a>$ (top panel), $<b/a>$ (middle panel)
 and triaxiality parameter $<\vartheta>$. The filled and open symbols represent
 simulations with barred orbits and circular orbits, respectively. The corresponding
 small symbols show the standard deviation, of these quantities.}
\label{fig:fig13}
\end{figure}

 In Fig.~\ref{fig:fig13} we plot the time-averaged values of the axial ratios
 $<c/a>$ and $<b/a>$, as well as of the triaxiality parameter
 $<\vartheta>$, all as a 
 function of the Jacobi integral $E_{\mathrm J}$ of the underlying periodic
 orbit. We find that star clusters at larger $E_{\mathrm J}$ roughly
 maintain their 
 spherical symmetry, much more so than clusters on orbits with small
 $E_{\mathrm J}$ (i.e. clusters near the centre), which become
 considerably triaxial. Clusters nearer to the centre also have
 larger dispersions of their axial ratios.  
 The {\em flattening} <c/a> of the cluster follows the same trend for
 simulations in barred and in axisymmetric  potentials. This is not
 true for the {\em elongation}
 <b/a> in the orbital plane, which has a stronger dependence on
 $E_{\mathrm J}$ in barred than in non-barred cases. These effects are
 presumably the reflections of the variation of the tidal force along
 the corresponding periodic orbit.

%------------------------------------------------
\subsection[]{Tidal forces}
\label{sec:tidal}
%------------------------------------------------

\begin{figure}
\begin{center}
\includegraphics[width=0.95\columnwidth]{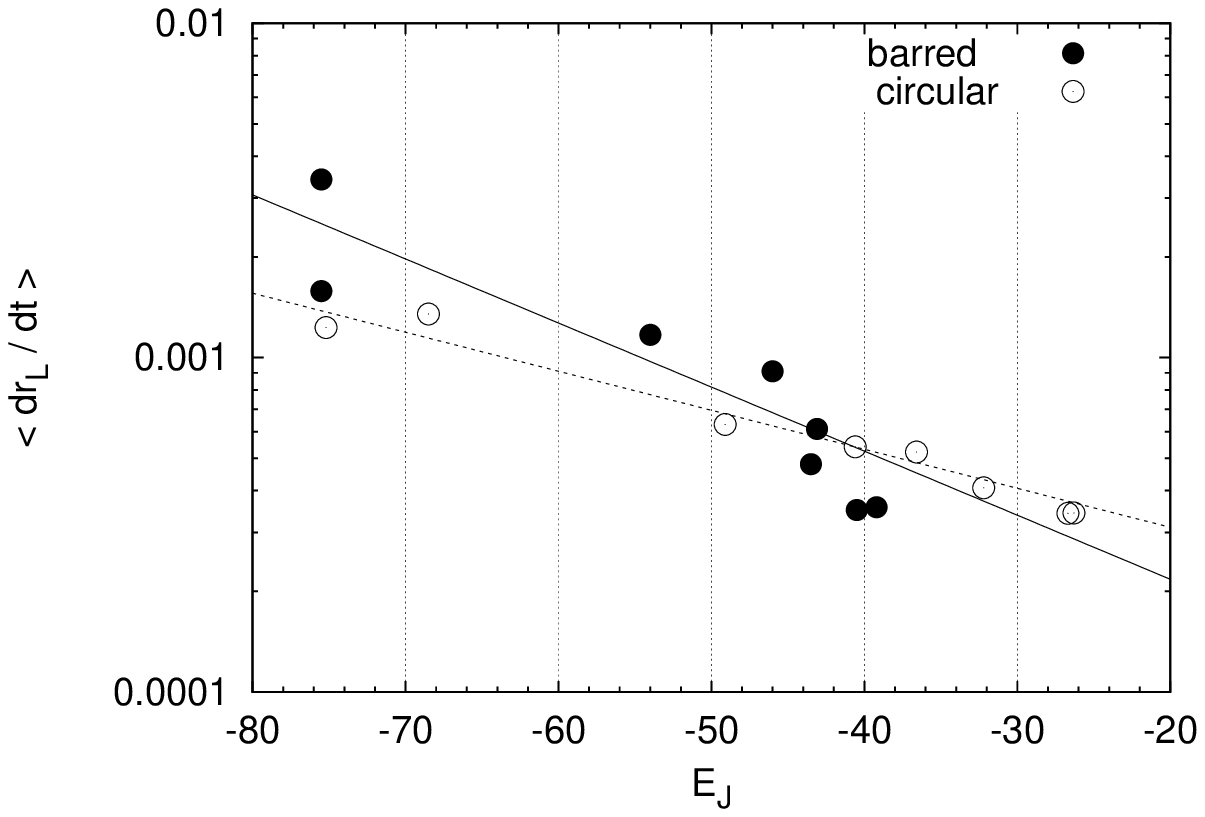}
\includegraphics[width=0.95\columnwidth]{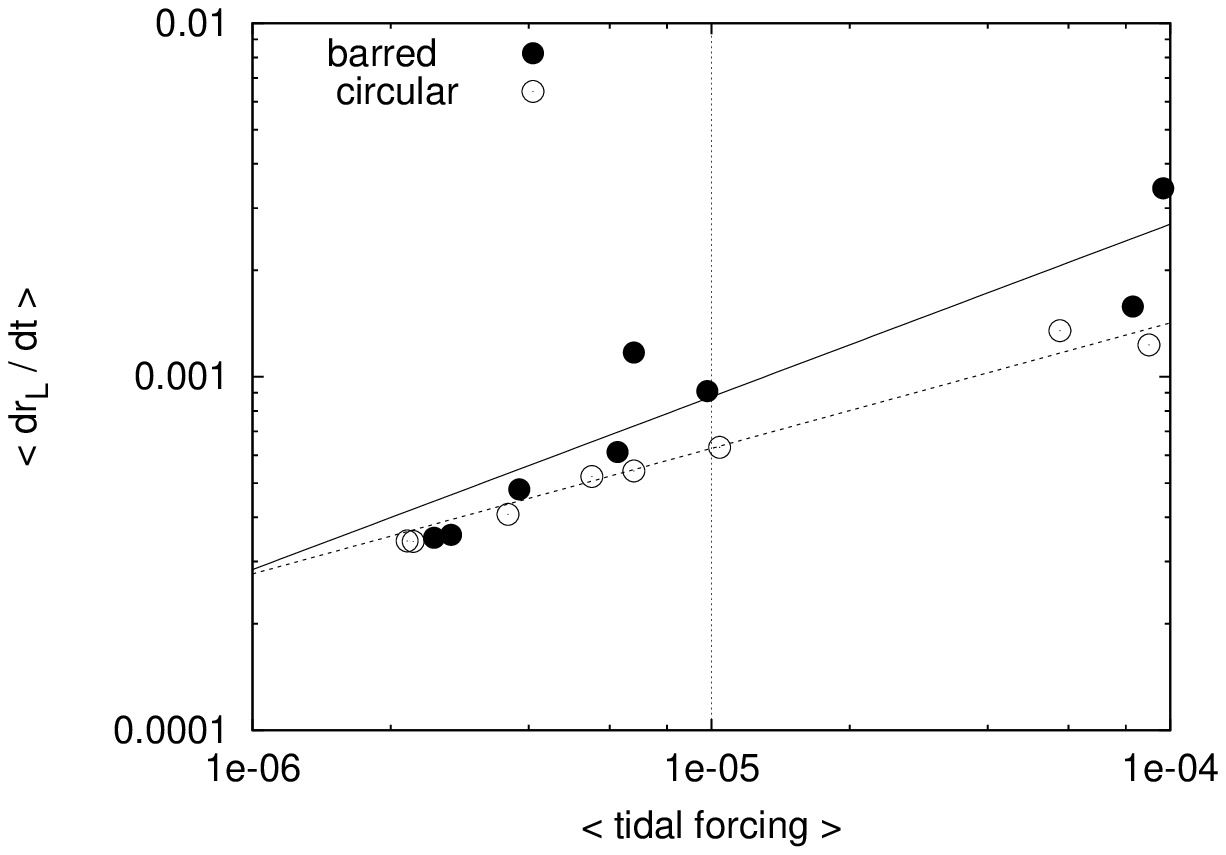}
\end{center}
\caption{Cluster expansion rate based on a linear fit to the (logarithmic)
 Lagrange radii of the outer mass shells. Filled and open symbols
 indicate simulations in barred and unbarred potentials, respectively.}
\label{fig:fig14}
\end{figure}

\begin{figure}
\begin{center}
\includegraphics[width=\columnwidth]{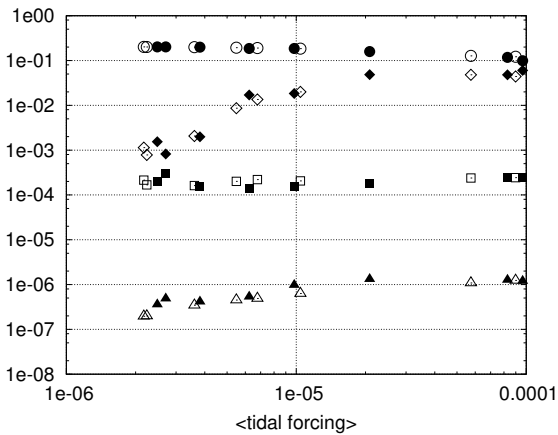}
\end{center}
\caption{Mass loss parameters as a function of the average tidal
 forcing. Filled and open symbols represent simulations in barred
 and axisymmetric potentials, respectively. The parameter of
 the linear fit $f_2(t)=m_2 - m_3\,t$ are plotted with circles
 ($m_2$) and triangles ($m_3$). The parameter of the exponential
 fit $f_1(t)=m_0 e^{-m_1\,t}$ are plotted with diamonds ($m_0$)
 and squares ($m_1$).}
\label{fig:fig15}
\end{figure}

\begin{figure}
\begin{center}
\includegraphics[width=0.95\columnwidth]{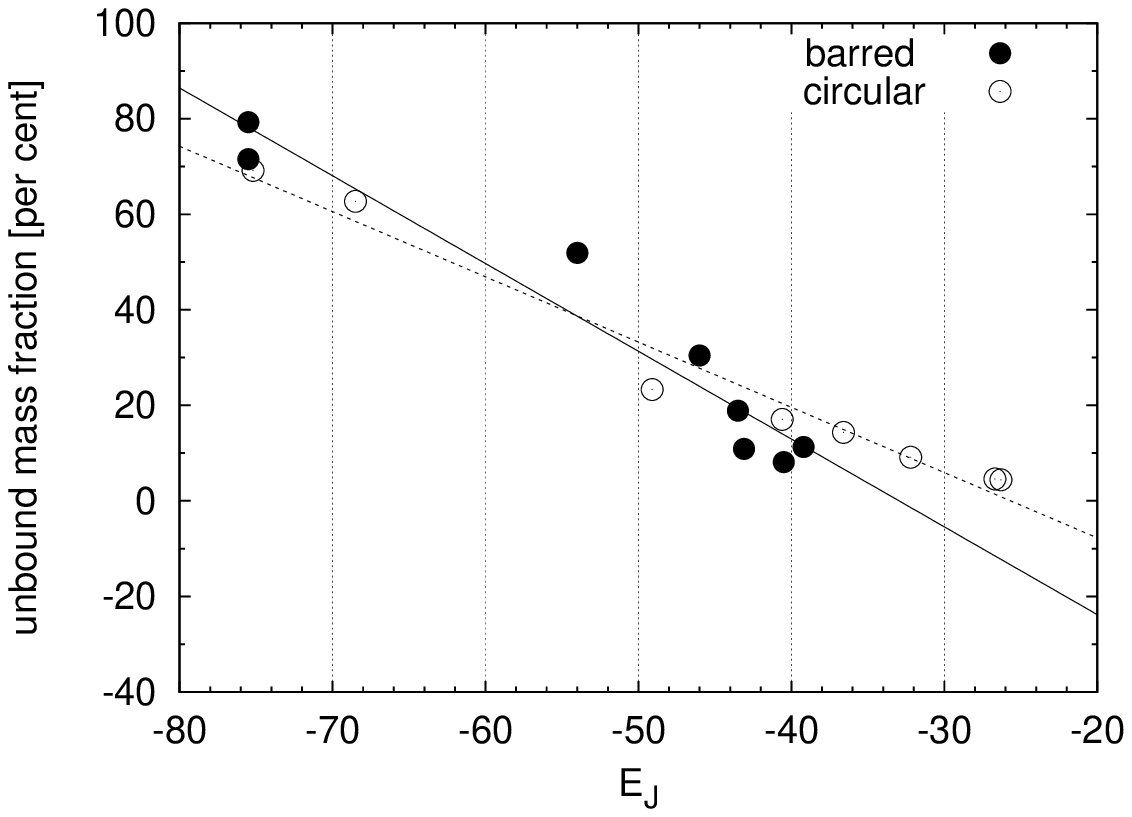}
\includegraphics[width=0.95\columnwidth]{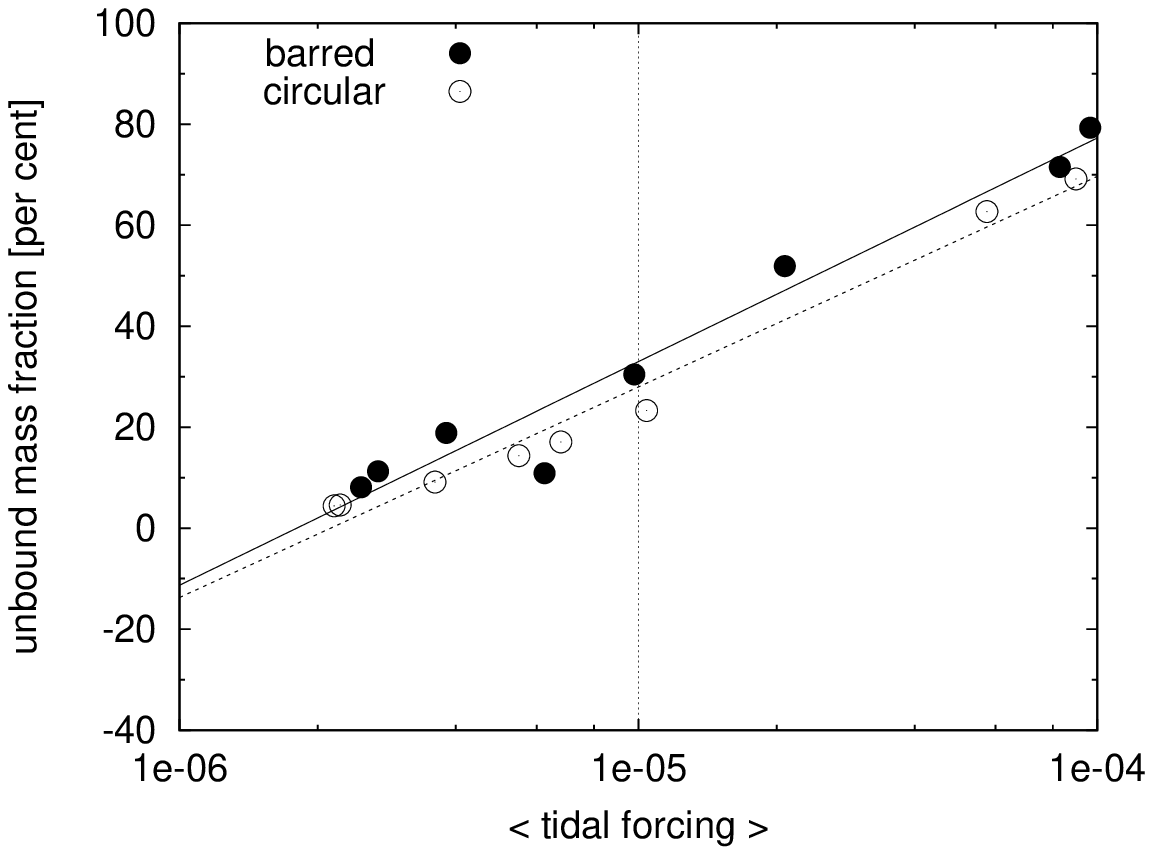}
\end{center}
\caption{Final mass loss (in per cent) in unbound material. Filled
 and open symbols indicate simulations in barred and unbarred
 potentials, respectively.}
\label{fig:fig16}
\end{figure}

 To quantify the strength of the tidal forcing along the periodic
 orbits we calculate the eigenvalues ($q_1, q_2, q_3$) and the
 corresponding eigenvectors of the force tensor\footnote{This
 definition is  similar to that of the tidal tensor used in,
 e.g., \cite{RBFNT08} and \cite{PC09}.}, as given by

\begin{equation}
 Q_{ij} = \frac{\partial ^2 \Phi_{\mathrm{gal}}}{\partial x^i \partial x^j}  \, .
\end{equation}
 
 We find that one of the eigenvectors is always parallel to the
 $z$-axis, i.e., oriented perpendicular to the orbital plane. In
 the following we call the corresponding eigenvalue $q_z$ and the
 two remaining eigenvalues $q_a$ and $q_b$. The eigenvectors of 
 $q_a$ and $q_b$ lie within the orbital plane. Positive eigenvalues
 are found for expansive flows, while negative ones represent for
 compressive flows. We measure the {\em tidal forcing} on the
 cluster by the quantity $\sqrt{q_a^2 + q_b^2}$. In Fig.\ref{fig:fig02}
 we use this quantity for the colour coding of the orbits. Note that
 the tidal forcing in barred potentials is strongest along the bars
 intermediate axis $b_{\mathrm{bar}}$ for our simulations with
 $E_{\mathrm J }\leq -43.1$ and along the bars major axis $a_{\mathrm{bar}}$
 otherwise.

\begin{figure*}
\begin{center}
\includegraphics[width=0.75\textwidth]{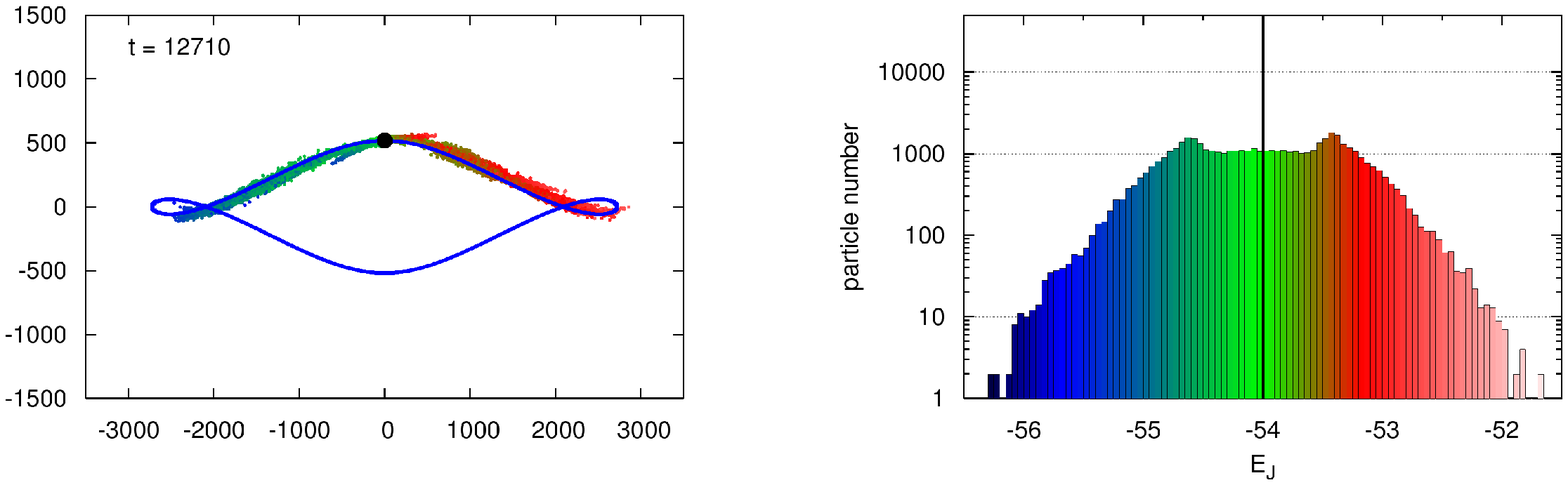}\\[-4ex]
\includegraphics[width=0.75\textwidth]{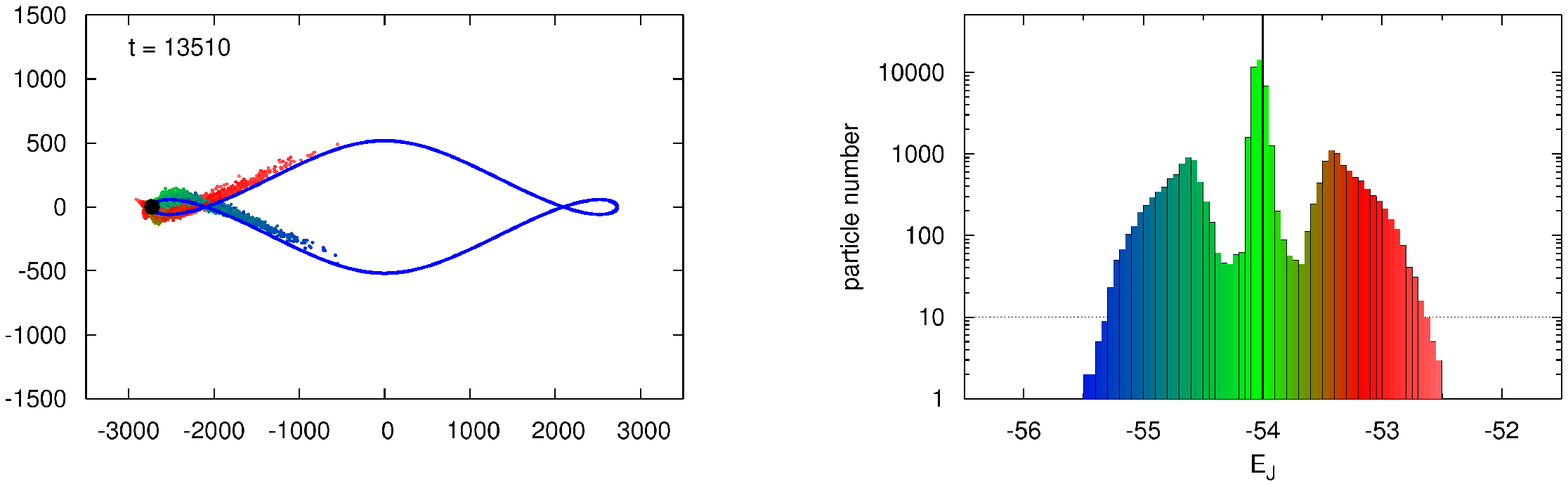}\\[-4ex]
\includegraphics[width=0.75\textwidth]{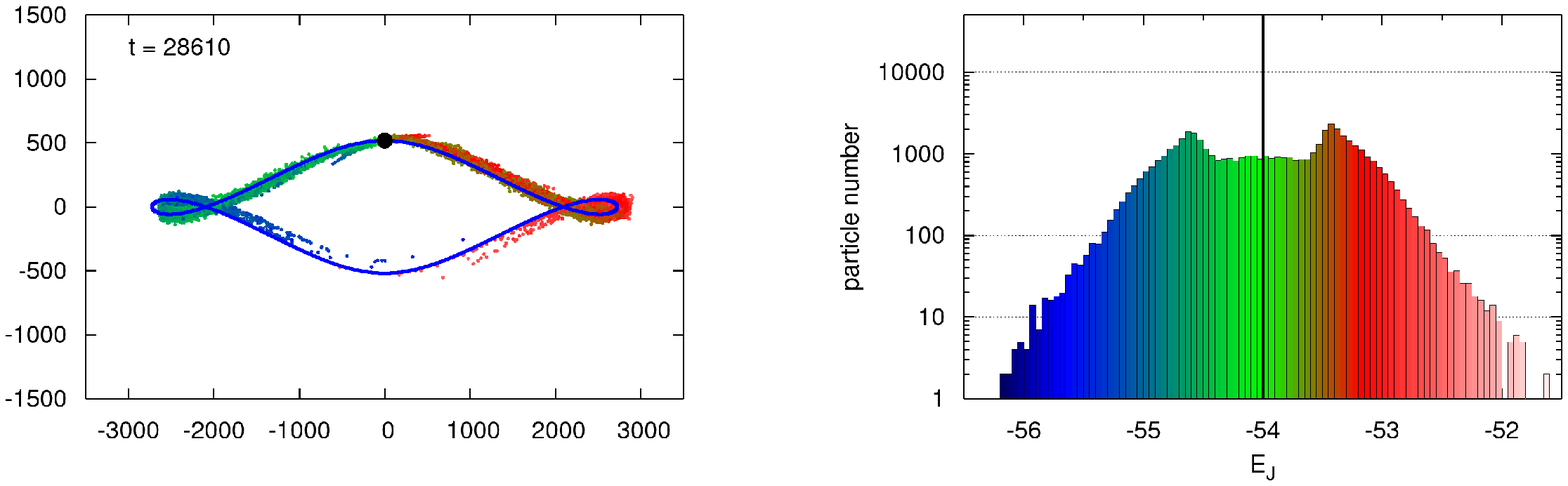}\\[-4ex]
\includegraphics[width=0.75\textwidth]{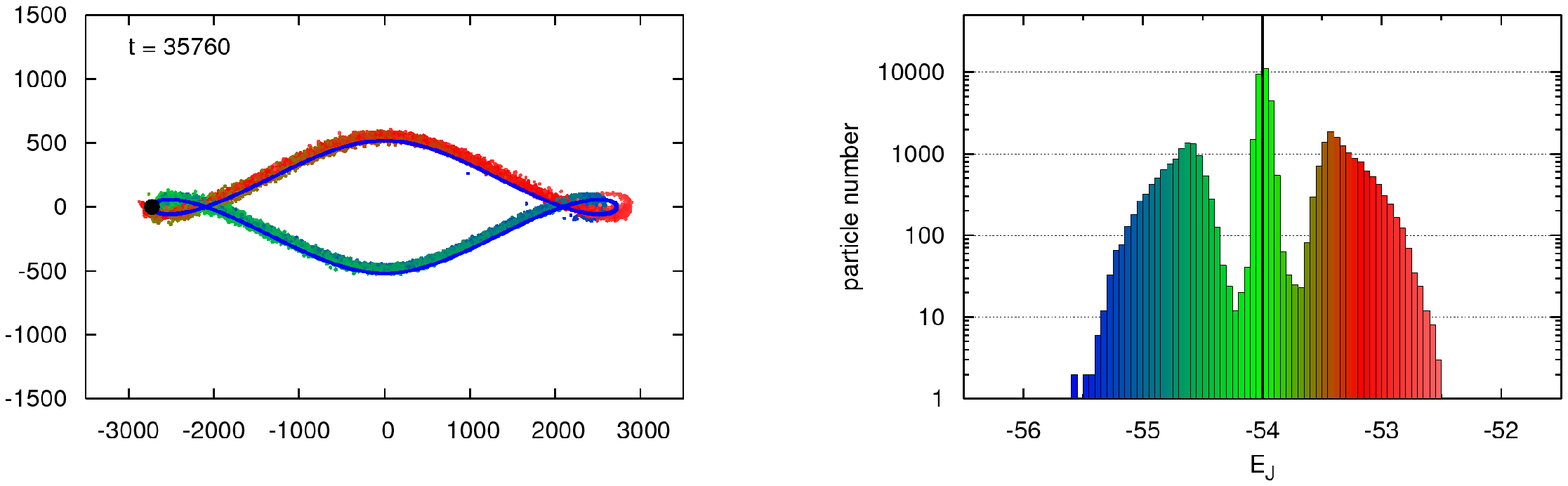}\\[-4ex]
\end{center}
\caption{Distribution of Jacobi energy $E_{\mathrm{J}}$ for simulation D1. The
 times from top to bottom correspond to 16, 16\,$^1\hspace{-0.75ex}/\hspace{-0.25ex}_4$,
 36, and 44\,$^1\hspace{-0.75ex}/\hspace{-0.25ex}_4$ orbital periods, respectively.
 The left panels show the projected particle distribution colour coded by $E_{\mathrm{J}}$. 
 The underlying periodic orbit is shown by the blue line and the position of the cluster
 is marked with a black dot. The right panels show the binned $E_{\mathrm{J}}$ distribution
 using the same colour coding as before. The Jacobi integral of the underlying
 periodic orbit is indicated by the vertical thick line.}
\label{fig:fig17}
\end{figure*} 

 As can be seen in the figures showing Lagrange radii, the expansion rate
 of the different mass shells typically seems to be higher for orbits
 in the  barred potential as compared to the circular ones.
 In Fig.~\ref{fig:fig14} we show the rate at which the  cluster is 
 expanding/dissolving. We determine this rate by using a linear fit to
 the steeply rising part of the Lagrange radii of the outer mass shells.
 We then calculate the average expansion rate $<dr_{\mathrm L}/dt>$ for
 each run from about $5$ determined slopes. The top panel in 
 Fig.~\ref{fig:fig14} shows the expansion rate as a function of the Jacobi
 integral $E_{\mathrm J}$. We find a strong correlation between the expansion rate
 and the Jacobi integral. Equivalently, we also plot the expansion rate
 as a function of the orbit-averaged\footnote{See Sec.~\ref{sec:SimA}
 for details.}
{\em tidal forcing} (see bottom panel in
 Fig.~\ref{fig:fig14}). Generally,
 the expansion rate is found to be higher for clusters in barred
 potentials (at least for small $E_{\mathrm J}$).

 In all our simulations the clusters lose their mass (i.e., stars
 become gravitationally unbound) roughly linearly. For clusters
 starting close to the centre of the galaxy, we observe an additional,
 exponentially decaying mass loss, as a consequence of the non-equilibrium
 initial conditions of the cluster in the strong tidal field of the galaxy. 
 To quantify the mass loss of the clusters in our simulations, we fit
 $f_{\mathrm{ml}}$ from our Eq.~\ref{eq:massloss} to the mass as a
 function of time curves, such as shown in Fig.~\ref{fig:fig07}. The
 results of these 
 fits is shown in Fig.~\ref{fig:fig15}. For simulations with a weak
 tidal forcing we get a $y$-intercept of the linear contribution
 of $m_2=0.2$, which corresponds to the initial cluster mass. This
 shows that, in
 this case, there is no significant contribution from the exponential
 mass loss. For higher values of the tidal forcing (and the initial
 tidal perturbation), 
 the exponential part $f_1(t)$ becomes more and more
 important. At the same time, the value of $m_2$ drops to roughly
 $0.1$. The linear mass loss rate ($m_3$), shown as triangles in
 Fig.~\ref{fig:fig15}, increases with increasing tidal forces. The
 solid triangles (barred models) lie slightly higher than the open
 ones (axisymmetric potential).

 The exponential mass loss rate $m_1$ (squares) is roughly independent
 of the tidal forcing, but the amplitude $m_0$ (diamonds) increases considerably
 with increasing tidal forcing. To summarise, it seems that if we compare
 clusters in barred and axisymmetric potentials that there is on average
 very little difference in the overall mass loss rate, if the average tidal
 forcing along the orbits is the same.

 In Fig.~\ref{fig:fig16} we show the total mass loss of
 the cluster determined at the end of the simulations.
 We again find strong correlations between this mass
 loss and the Jacobi integrals of the orbits or
 the average tidal forcing.

%------------------------------------------------
\subsection[]{Energy considerations}
\label{sec:energy}
%------------------------------------------------

 In Figures~\ref{fig:fig17}
 we show the evolution
 of the energy distribution within the cluster for simulations D1
 (barred potential).
 This is of particular interest concerning the interpretation of 
 observational data on the virial state of a star cluster \citep{KMK11},
 e.g., such as Palomar 13.
  
 Fig.~\ref{fig:fig17} shows the  particle distribution in simulation D1
 (left panels) and the corresponding energy distribution (right panels),
 both with a colour scheme based on their Jacobi integral. The snapshots
 are chosen at times when the cluster is at apo- (second and fourth
 rows) and peri-centre (first and third rows). With
 the formation of the tidal tails and the accompanied dissolution of the
 main cluster body the energy distribution slowly forms a double peak
 structure roughly 
 symmetric with respect to the Jacobi integral of the periodic orbit.
 Thus the whole distribution has a three peak structure: the two peaks
 corresponding to the tidal tails, plus the central peak of the initial
 cluster distribution. The latter can at times be relatively low, so that
 the central structure may look like a plateau, particularly with the
 logarithmic scale we are using. At pericentres the double peaks reach
 higher values than at apocentres. However, the relative contrast between
 the peaks becomes much strongest at apocentres, so that they stand out
 clearest. This is due to the compression effect described in
 Sect.~\ref{subsub:C1C2}. 
 Note that the central peak is not pronounced at pericentres, but stands
 out clearly at apocentres, where it is an order of magnitude higher than at
 pericentre. Comparing these structures always at the same phase of the orbit
 (i.e. always at apocentre, or always at pericentre) we find that, as time
 increases, the height of the two side peaks increases, that of the central
 peak decreases and the contrast between the peaks becomes much sharper.
 This is due to the fact that more and more material leaves the cluster
 and moves to the tails, as seen in Sect.~\ref{sec:results}. 

 Note that the double peak structure mentioned above is also visible in
 simulations with circular orbits, but is nevertheless much less pronounced
 - due to the absence of the compression/expansion along the circular
 orbit. Further discussion of the tails and of the corresponding $E_{\mathrm J}$
 distributions will be given elsewhere (Athanassoula, R\'omero-Gomez and Berentzen,
 in prep.).

%-------------------------------------------------------------------------------------------------------------------
\section{Summary and conclusions} \label{sec:conclusions}
%-------------------------------------------------------------------------------------------------------------------

 Stellar streams are fossils of the formation history of galaxies
 and can be used to reconstruct, e.g., the shape of the gravitational
 potential of dark matter halos. Several issues such as the life-time
 of the clusters and their tidal streams however are not yet fully
 understood. The understanding of the morphology and the dynamics
 of the tails and their sub-structures is still in its infancy.
 Numerical $N$-body simulations are a main key to address these 
 questions. Many simulations in the literature focus on the evolution
 of dense stellar systems in an external axisymmetric potential on
 both circular and eccentric orbits, respectively. For orbits close
 to the galactic plane, however, non-axisymmetric perturbations in
 the disk, such as stellar bars and/or spiral arms, are expected to
 significantly influence the dynamical evolution
 of star clusters.
   
 In this work we have presented a set of direct $N$-body simulations
 of star clusters orbiting in the plane of a barred galaxy potential.
 The clusters are placed in the mid-plane of the galactic disc and
 launched on periodic orbits which have been selected from the main
 planar 2-d families. The results are then compared to the evolution
 of star clusters on circular orbits in an axisymmetric potential.
 
 We find that the star clusters dissolve due to the tidal force field
 of the galaxy -- in both the axisymmetric as well as in the barred
 potential. While the tidal tails in the axisymmetric potential 
 continuously grow in length, we find a periodic compression and
 expansion of the tails for star clusters in the barred potential.
 Both the length and the density of the tidal tails vary with time
 and are directly correlated to orbital period of the underlying
 parental orbit. We find sub-structures within the tidal tails in
 form of clumps (or epicyclic over-densities). The qualitative evolution of these
 structures is in agreement with the results of \citet{KKBH10},
 who studied the evolution of star clusters on eccentric orbits
 in a static axisymmetric potential.

 We also studied the distribution of the Jacobi energy $E_{\mathrm J}$ of the
 cluster stars and found that it depends strongly on the location of the cluster
 along the orbit. It remains, nevertheless, roughly symmetrical with respect
 to the Jacobi energy of the cluster orbit. Two peaks develop, one on each side
 of this energy, and their relative amplitude depends on the location of the
 cluster along the orbit. At apocentres the distribution is tri-modal and the three peaks
 are very clearly separated. At pericentres, the distribution is much
 more extended and the central peak is not clearly delineated. 

 We find that the mass loss of the cluster is mainly determined by the
 orbit-averaged tidal forcing. In other words, the shape of the
 gravitational potential -- at least in the case studied here --
 affects the dissolution time-scale only very little.
 On the other hand, the stellar bar has a strong effect on the
 shape and morphological evolution of the tidal tails: the 
 stability of the orbit has a strong impact on how long
 the tidal tails and their sub-structures actually survive.
 This makes it more subtle to reconstruct the cluster orbits
 in observations and thus to probe that galactic potential
 in the disc plane. This could also be the case small
 galactic altitudes in which the 3-d shape of orbits is
 strongly affected by the stellar bar
 \citep[e.g.,][]{Pfe84, BHSF98, SPA02a, SPA02b, PSA02}.

%======================================================
%======================================================

\section*{Acknowledgments}
 We would like to thank the anonymous referee for her/his comments
 and suggestions which helped to improve the manuscript. We also would
 like to thank Rainer Spurzem, Andreas Just, Peter Berczik, Andreas
 Ernst and Albert Bosma for many fruitful discussions. We
 are grateful to Jean-Charles Lambert for his most valuable help
 and assistance. IB acknowledges the kind hospitality and the support
 from the Observatory Astronomique de Marseille-Provence during his
 visit in which this work has been initiated. He also acknowledges
 funding through the University of Marseille for his visit. Furthermore,
 IB acknowledges funding through a Frontier Innovation grant of the
 University of Heidelberg sponsored by the German Excellence Initiative,
 as well as funding by the Sonderforschungsbereich SFB 881 ``The Milky
 Way System'' (subproject A1) of the German Research Foundation (DFG).
 The Kolob GPU Cluster used for most simulations in this work is 
 funded in part by the DFG via Emmy-Noether grant BA 3706.

%==========================================================================================
%
%==========================================================================================

\bsp

\label{lastpage}
\end{document}